\documentclass[aps, prb, reprint]{revtex4-2}

\usepackage{epstopdf}
\usepackage[caption=false]{subfig}

\usepackage{amsmath}
\usepackage{amssymb}
\usepackage{graphicx}

\usepackage{hyperref}
\usepackage{xcolor}

\begin{document}


\title{On the chemical potential and grand potential density of solids \\ under non-hydrostatic stress}


\author{Michiel Sprik}
\email{ms284@cam.ac.uk}
\affiliation{Yusuf Hamied Department of Chemistry, University of Cambridge, Lensfield Road, Cambridge CB2 1EW, United Kingdom}


\begin{abstract}
Non-hydrostatic stress has a peculiar effect on the phase equilibrium between solids and liquids.  This was already pointed out by Gibbs. Gibbs derived his formulation of the condition for liquid-solid coexistence applying a surface accretion process without imposing chemical equilibrium between liquid and solid.  Adding particles to the bulk of a solid was not possible in his view at the time. Chemical potentials for solids were later introduced by material scientists. This required extending chemical and mechanical equilibrium with a third condition involving a relation between grand potential densities controlling the migration of the interface. These issues are investigated using a non-linear elastic continuum model (technically an open compressible neo-Hookean material) developed in a previous publication (M.~Sprik, J.~Chem.~Phys.~{\bf 155}, 244701 (2021)). In common with a liquid, the grand potential density of the model is equal to minus the mean pressure even if the stress is non-hydrostatic. Applying isothermal compression normal to a liquid-solid interface initially in hydrostatic equilibrium drives the system away from coexistence.  We derive the Gibbs-Thomson correction to the pressure of the liquid required to restore phase equilibrium. We find that the coupling between chemical potential of the solid and shear stress is a purely non-linear effect.  
\end{abstract}


\maketitle

\section{Introduction}\label{intro}

The thermodynamic conditions for phase equilibrium between a liquid and gas are fundamental in physical chemistry. In equilibrium at a given temperature, a one-component fluid  and its vapour must have the same pressure and chemical potential. These rules were established by Gibbs and are a highlight  in every text book on molecular thermodynamics. Gibbs equally derived conditions for  phase equilibrium between a solid and its melt or with its constituent components dissolved in a liquid. Here Gibbs followed an approach which can come as somewhat of a surprise to a modern physical chemist: Gibbs denied a solid a chemical potential. To Gibbs a solid was a compact piece of matter, somewhat like a macroscopic molecule. Vacancies and diffusion in solids were not known at the time of Gibbs and he assumed that it is not possible to insert atoms.   Atoms can only be added at the surface of a solid. This process, usually referred to as accretion in material science, replaces  insertion as the fundamental process for mass increase of a solid. The equilibrium condition for accretion is derived by Gibbs without coupling to a chemical potential\cite{gibbscw}. 

Soon after material science demonstrated the existence of mobile vacancies and the Gibbs accretion process was extended to include chemical potentials for solids\cite{cahn73,cahn78,cahn80,cahn85,mullins84,mullins85,sekerka89,voorhees04,cammarata09}.  Chemical equilibrium at coexistence again requires the chemical potential of the  solid  and liquid to be equal consistent with the thermodynamics for liquid-vapour phase equilibrium.   However, essential differences remain. The combination of chemical and mechanical equilibrium is not a sufficient condition for coexistence of a solid with its melt. An additional phase relation is needed to eliminate the driving force for the migration of the solid-liquid interface. This condition involves the grand potential density. Formally a similar condition applies at a liquid-vapour interface but for liquids  the grand potential density is identical to (minus) the pressure making this condition redundant.  This is not always the case for a solid-liquid interface.  Discrepancies arise for anisotropically stressed solids.  Non-hydrostatic stress is a confusing complication for the thermodynamics of solids as was already noted by Gibbs\cite{gibbscw,cahn80}.  He captured this observation in his famous prism paradox stating that the chemical potential of the solid component in solution is different depending on the face of the solid body exposed to the liquid. The problem persists even disregarding the question of the specification of a chemical potential for a solid. Once dissolved the solid particles are a regular thermochemical species with a well defined chemical potential.

The Gibbs prism paradox is an extreme expression of the distinction between the thermodynamics of (uniform) liquids and solids. Allowing for interface energies (surface tension and stress) leads to further complications as also already noted by Gibbs himself\cite{gibbscw,cahn80}.  Solid thermodynamics  has been worked out in a modern fomulation  by Cahn, Herring, Mullins, Sekerka and others applying  the Gibbs variational principle to the thermodynamic variables of a crystal\cite{cahn73,cahn78,cahn80,cahn85,mullins84,mullins85,sekerka89,sekerka04,voorhees04,cammarata09}.  A synthesis  accounting for the combination of mobile interfaces, strain and vacancies is given Ref.~\citenum{sekerka89}.   This work also formally identifies the grand potential density gap as the driving force for interface migration (assuming the process is strictly isothermal).  In absence of interface energies the grand potential densities of two phases in  equilibrium must be equal. This relation imposes a third thermodynamic constraint supplementing the balance of the normal components of stress  (mechanical equilibrium) and equal chemical potentials (chemical equilibrium).  

\newpage

The importance of grand potential density gaps for the stability of dividing surfaces  is well recognized in the material science literature. It is less familiar to physical chemists. The author was certainly unaware of this issue in his study of the thermomechanics of melting of an elastic solid published some time ago\cite{sprik21c}. This work was based on a non-linear elastic model as used in the continuum mechanics of finite deformations\cite{ogden97,gurtin10}. The model is defined by specifying a Lagrangian reference free energy density.  The free energy density is the sum of a  shear elastic energy term and a term controlling isotropic change of volume similar to a compressible Neo-Hookean model for rubber\cite{ogden97}. The isotropic term has the form of a non-linear function of the actual (Eulerian) particle density as is characteristic for a simple liquid.  With this modification the solid can be transformed into its liquid phase by simply switching off the shear elastic modulus without having to worry about a reference for the bulk modulus of the liquid.

The stress tensor is determined by evaluating the derivative of the reference free energy density with respect to the deformation gradient  (Piola stress tensor).  This is standard procedure in non-linear elasticity.   The chemical potential is the complementary  material derivative.  The energy density is differentiated to the particle density in the reference frame,  treating the density prior to elastic deformation as an additional thermodynamic degree of freedom.   Using  balance laws from irreversible non-equilibrium thermodynamics  Gurtin, Fried and Anand showed that this reference (material) particle derivative is the unique chemical potential which can be transferred to  deformed system without change.\cite{gurtin10}.  This fundamental property of the chemical potential also follows from the theory in Ref.~\citenum{sekerka89} using the Gibbs variational principle.  The grand potential density is then evaluated substituting the chemical potential in the usual expression given by Gibbs.  

The  grand potential density generated by our model deviates from  what is generally regarded as characteristic for a solid.  The grand potential density of a liquid can be interpreted as a thermodynamic pressure.  For solids the thermodynamic pressure is normally  different from the mechanical pressure defined as minus a third of the trace of the Cauchy stress tensor.  However, for the model introduced in Ref.~\citenum{sprik21c} these two pressures are equal even under non-hydrostatic stress.  The solid retains some liquid-like character in response to volumetric elastic deformation.  A first aim of the present work is therefore providing a proof of this property which will turn out to be rather convenient in thermodynamic derivations. 

 The model was applied in Ref.~\citenum{sprik21c} to explore the stability under non-hydrostatic stress of a block of solid in phase equilibrium with surrounding liquid.  Surface energies and vacancy formation were ignored.  The solid was subjected to uni-axial compression while constraining dilation in the two perpendicular directions. The conclusion was that the compressed solid remains stable.  Melting is arrested by the build-up of shear stress. Resistance of an open system to non-hydrostatic stress is observed  for a gel but is rather unexpected for a pure one-component solid. This somewhat counterintuitive result was obtained by imposing mechanical equilibrium (continuity of stress normal to an interface) and chemical equilibrium (equal chemical potentials) but disregarding the zero grand potential gap condition.  This simple liquid-solid phase equilibrium problem is revisited in the present work but now also imposing continuity of the  grand potential density.  

Applying this extended thermomechanical scheme we find that chemical equilibrium and uniform thermodynamic pressure are incompatible for a solid driven away from hydrostatic equilibrium with its liquid phase by the application of non-hydrostatic stress (again ignoring interface excess energies and defects).   Under these thermodynamic boundary conditions the solid-melt interface is not stable contrary to what was stated in our previous publication\cite{sprik21c}.  Phase equilibrium can however be restored by adjusting the pressure of the liquid.  This correction is an example of a Gibbs-Thomson (GT) relation and will be determined for the open system non-linear elastic model taken over from Ref.~\citenum{sprik21c}.  What is added by the non-linearity is coupling between the chemical potential and shear strain.   Linearization of the elastic response reduces the non-linear GT pressure correction to the Sekerka-Cahn type expression\cite{sekerka04} used in the material\cite{mishin10}  and earth science literature\cite{speck24}. This limiting behaviour required by consistency will be verified following Nozi\'{e}res in his elementary treatment of GT relations in isotropic systems\cite{nozieres93,nozieres95}.

The outline of the paper is follows.  In section \ref{section:Gibbs} the balance equation for the Gibbs accretion process is summarized and  applied in a heuristic derivation of the zero grand potential potential gap condition (A rigorous derivation can be found in Refs.~\citenum{mullins85} and \citenum{sekerka89}. See also the review by Voorhees and Johnson \cite{voorhees04}).  The non-linear elastic model of Ref.~\citenum{sprik21c} is introduced in section \ref{section:nlelast} followed by the determination of the thermodynamic derivatives in section \ref{section:soltherm}.   Sections \ref{section:nlelast} and \ref{section:soltherm} are a review of the technical material in Ref.~\citenum{sprik21c} now also accounting for the grand potential density.  In section \ref{section:application} the thermodynamic stability conditions for a melting solid under non-hydrostatic stress are revisited applying the non-linear model of section \ref{section:nlelast} including the determination of the isothermal Gibbs-Thomson pressure correction. The results are analyzed in section \ref{section:discussion}.  Section \ref{section:conforce} is a (preliminary) attempt to place the results for the model system in the wider context of configurational mechanics\cite{gurtin10,gurtin95,gurtin99,maugin93,gurtin00,kienzler00,maugin11}. The main point will be that the grand potential density gap can be considered as an example of a ``grand'' configurational force acting on an internal dividing surface under chemical equilibrium conditions.  Limitations of the continuum model are discussed in sections \ref{section:limitations}  leading to some suggestions for the improvements, in particular  a very preliminary account of the effect of vacancies.    We  conclude with a recapitulation and outlook in section \ref{section:conclusion}.
  
\section{Surface accretion}\label{section:Gibbs}

\subsection{Phase equilibrium without chemical equilibrium} \label{section:nomu}
The thermodynamic scheme for melting of a solid body devised by Gibbs consists of removing a thin layer of material from its surface and replacing it by liquid.  The amount of liquid substituting for the solid is the same as the amount of solid removed.  For a one-component system this amount can be specified in terms of the number of moles $\Delta m$.  The process is isothermal and the cost in free energy for the loss of solid can therefore be computed from the molar Helmholtz free energy $\psi_m^{\mathrm{S}} $ 
\begin{equation}
\Delta F^{\mathrm{S}} = - \psi_m^{\mathrm{S}}  \Delta m 
\end{equation}
 Superscripts denote the phase of the subsystem. The result is a strip of empty space separating the solid phase S and liquid phase L.  The liquid plays the role of a reservoir with a fixed hydrostatic  pressure $p^{\mathrm{L}}$.  
 
 A liquid has in general a larger  molar volume $v_m$ than its solid phase S: $v_m^{\mathrm{L}}  > v_m^{\mathrm{S}}$.  Therefore before  filling up the strip vacated by the solid  with liquid at the same pressure $p^{\mathrm{L}}$  as the reservoir  its volume  must be adjusted by pushing back the  interface with the bulk liquid reservoir. The requires an amount of mechanical work
 \begin{equation}
W = - p^{\mathrm{L}} \Delta V 
\end{equation}
with the change in volume $\Delta V$ given by 
\begin{equation}
 \Delta V = - \left(v_m^{\mathrm{L}}  - v_m^{\mathrm{S}} \right )\Delta m 
\end{equation}
As we have assumed that  $v_m^{\mathrm{L}}  > v_m^{\mathrm{S}}$,  the work $\Delta W$  is positive  because $\Delta V$ is not the change in volume of the interface layer but of the liquid reservoir. Now we are ready to insert the liquid gaining a free energy of 
 \begin{equation}
\Delta F^{\mathrm{L}} =  \psi_m^{\mathrm{L}}  \Delta m 
\end{equation}
Maintaining thermal equilibrium the net free energy change vanishes
\begin{equation}
\Delta F^{\mathrm{S}} +  W + \Delta F^{\mathrm{L}} = 0
\end{equation}
Substituting and dividing out $\Delta m$ we find
\begin{equation}
  - \psi_m^{\mathrm{S}}  +   p^{\mathrm{L}}  \left(v_m^{\mathrm{L}}  - v_m^{\mathrm{S}} \right ) + \psi_m^{\mathrm{L}} = 0
\end{equation}
which becomes 
\begin{equation}
\psi_m^{\mathrm{S}} + p^{\mathrm{L}}  v_m^{\mathrm{S}} = \psi_m^{\mathrm{L}}   + p^{\mathrm{L}}  v_m^{\mathrm{L}}
\label{Gibbsaccr}
\end{equation}
after rearranging

Eq.~\ref{Gibbsaccr}  will be referred to as the Gibbs accretion condition\cite{gibbscw,cahn80,sekerka04,cammarata09}. The right hand side (rhs)  is the molar free enthalpy of the liquid phase and can be safely exchanged for the corresponding chemical potential $\mu^{\mathrm{L}}$ 
\begin{equation}
 \psi_m^{\mathrm{S}}  + p^{\mathrm{L}}  v_m^{\mathrm{S}} = \mu^{\mathrm{L}}
 \label{mulaccr}
\end{equation}
In case of accretion of solid in hydrostatic equilibrium with the bulk liquid,  the solid has a well defined pressure  $ p^{\mathrm{S}} =  p^{\mathrm{L}}$. The left hand side (lhs) of Eq.~\ref{mulaccr} can then be identified with a chemical potential $\mu^{\mathrm{S}}$ for the solid  recovering the familiar chemical equilibrium equation $\mu^{\mathrm{S}}=\mu^{\mathrm{L}}$.  An important point to note is that equality of chemical potentials is not an  apriori assumption but an implication of accretion under hydrostatic pressure.  The argument is the other way around from the usual presentation in physical chemical textbooks.

Now consider a rectangular piece of solid (a prism) subjected to surface forces perpendicular to its faces. The surface forces  are different for the three  directions which will be labelled by index $i=1,2,3$.  The corresponding non-hydrostatic stress tensor is diagonal in an axis frame aligned along the edges of the prism. 
\begin{equation}
\boldsymbol{\sigma} = \begin{pmatrix} \sigma_{1} & 0 & 0 \\[4pt] 0 & \sigma_{2} & 0 \\[4pt]  0&  0&  \sigma_{3} \end{pmatrix}
\label{prismstress}
\end{equation}
The tractions (surface force densities) are directed along the surface normals $\mathbf{n}_i$ 
\begin{equation}
\boldsymbol{\sigma} \cdot \mathbf{n}_i = \sigma_i \mathbf{n}_i 
\label{prismforce}
\end{equation}
In mechanical equilibrium the stress Eq.~\ref{prismforce} is balanced by the force  exerted by the liquid on face $i$ (Cauchy equation).  By convention the pressure tensor is minus the stress tensor we must therefore have 
\begin{equation}
\sigma_i = -p^{\mathrm{L}}
\label{pgap}
\end{equation}
 This cannot be a mechanically stable state if $p^{\mathrm{L}}$ is to be the unique hydrostatic pressure  of a liquid surrounding the solid body unless the surface stress is also the same for each face.
 
From the perspective of solid mechanics, stability issues with non-hydrostatic stress such as Eq.~\ref{pgap} are perfectly normal.  What troubled Gibbs is the implication for phase equilibrium.  Inserting Eq.~\ref{pgap} in  his accretion condition Eq.~\ref{mulaccr} leads to
 \begin{equation}
 \psi_m^{\mathrm{S}} - \sigma _i v_m^{\mathrm{S}} =  \mu^{\mathrm{L}}
 \label{muparadox}
\end{equation}
which seems incompatible with maintaining chemical equilibrium with a liquid in which the prism is immersed. The conclusion of Gibbs was that the chemical potential of a species which is the component in solution of a solid in phase equilibrium with the solution depends on the surface exposed to the liquid.  If the solid is under non-hydrostatic stress the chemical potential varies with the surface orientation. This argument has become known as the Gibbs prism paradox.  However, a system in this state is not unphysical\cite{cahn80}.  It becomes only  problematic if we insist on assigning a chemical potential to the solid. In fact,  this has led some theorists to go as far as to declaring the chemical potential of a solid to be tensorial\cite{rusanov05}.

\subsection{Imposing chemical equilibrium} \label{section:withmu}
Gibbs did not regard the thermodynamic paradox bearing his name  immediately fatal  because he had his doubts about the validity  of chemical potentials for solids. The concept of a chemical potential for a solid (or a crystal to more precise \cite{cahn85}) is however fully accepted in modern statistical mechanics and supported by experiment\cite{weissmueller18}. Let us therefore try to reformulate the  argument of section \ref{section:nomu}  by  postulating  right from the start that accretion proceeds under chemical equilibrium and worry about mechanical equilibrium afterwards.  Hence assuming  that the usual relation for the chemical potentials at phase equilibrium applies
\begin{equation}
 \mu^{\mathrm{S}} = \mu^{\mathrm{L}}
 \label{mugap}
\end{equation}
we go back to the accretion condition Eq.~\ref{mulaccr}  before force balance  (Eq.~\ref{pgap})  was imposed 
and replace $\mu^{\mathrm{L}}$ by $\mu^{\mathrm{S}}$
\begin{equation}
\psi_m^{\mathrm{S}} + p^{\mathrm{L}}  v_m^{\mathrm{S}} = \mu^{\mathrm{S}}
\end{equation}
Dividing by $v_m^{\mathrm{S}}$ and recalling  that $ 1 /v_m^{\mathrm{S}} = \rho^{\mathrm{S}}$ is the density of the solid we can write
\begin{equation}
\psi^{\mathrm{S}} - \mu^{\mathrm{S}} \rho^{\mathrm{S}}= - p^{\mathrm{L}} 
\label{musaccr}
\end{equation}
where 
\begin{equation}
 \psi^{\mathrm{S}}  = \frac{\psi_m^{\mathrm{S}}}{v_m^{\mathrm{S}}} 
\end{equation}
is the Helmholtz free energy density of the solid. Having decided that $\mu^{\mathrm{S}}$ is the chemical potential of the solid, the lhs of Eq.~\ref{musaccr} can be identified with its grand potential density of the solid
 \begin{equation}
 \omega^{\mathrm{S}} = \psi^{\mathrm{S}} - \mu^{\mathrm{S}} \rho^{\mathrm{S}}
 \label{omegas}
 \end{equation}
 The rhs of Eq.~\ref{musaccr} is as always the grand potential density $\omega^{\mathrm{L}}$ of the liquid. Substituting, Eq.~\ref{musaccr} becomes an equilibrium condition for grand potentials
\begin{equation}
\omega^{\mathrm{S}} = \omega^{\mathrm{L}}
\label{omegap}
\end{equation}
Solid-liquid phase equilibrium must satisfy the  balance of grand potential densities Eq.~\ref{omegap} in addition to conditions Eqs.~\ref{pgap} and \ref{mugap} for stress and the chemical potential. 

The questions raised in section \ref{section:nomu} concerning the status of the chemical potentials of a solid are not resolved by simply adopting chemical equilibrium with the liquid as we have done in Eq.~\ref{mugap}. The Gibbs prism paradox is passed on to the grand potential density.  Substituting mechanical equilibrium Eq.~\ref{pgap} in Eq.~\ref{omegap}  converting  $\omega^{\mathrm{L}}$ back to $-p^{\mathrm{L}}$ gives
\begin{equation}
\omega^{\mathrm{S}} = \sigma_i
\label{omegaparadox}
\end{equation}
Again for a non-hydrostatically stressed solid there is a clash when faces with a different orientation are trying to reach phase equilibrium with the liquid.   Eq.~\ref{omegaparadox} is as troublesome as it was for the definition of the chemical potential in section \ref{section:nomu}. 

Fortunately, the Gibbs paradox  is not directly in the way of the application in this paper. Contact with the liquid is limited to only a single face (canister geometry). Eq.~\ref{omegap} is however crucial justifying a more rigorous derivation than the heuristic argument given in this section.  Here we rely on the fundamental paper by Leo and Sekerka\cite{sekerka89} which also clarifies the physical significance of a grand potential discontinuity.   As was recognized in material science following Gibbs, mechanical deformation is not the only degree of free freedom for an internal surface separating two phases. The interface is mobile and migration must be treated as a separate configurational degree of freedom (see also section \ref{section:conforce}).   Starting from this premise Leo and Sekerka show using the Gibbs variational principle that  the grand potential density gap is the conjugate thermodynamic force for this process provided it occurs under thermal and chemical equilibrium.

\section{Non-linear elastic model}\label{section:nlelast}

\subsection{Free energy density}\label{frendens}
The free energy of a liquid is specified by a free energy density $\psi^{\mathrm{L}}$ integrated over the volume $V$ occupied by the liquid.
\begin{equation}
\mathcal{F} = \int_V \psi^{\mathrm{L}} d v
\label{intFL}
\end{equation}
For a uniform liquid $\psi^{\mathrm{L}}$ is just a function of density $\rho = N/V$ where $N$ is the number of particles. In continuum mechanics this property is normally formulated as
\begin{equation}
\psi^{\mathrm{L}} = \hat{\psi}^{\mathrm{L}}(\rho)
\label{psiL}
\end{equation}
which is a formal constitutive statement of the nature of a simple fluid.  Non-simple fluids with an additional dependence on the gradient of density ($\psi^{\mathrm{L}} = \hat{\psi}^{\mathrm{L}}(\rho, \nabla \rho)$) will not be considered here.

For a continuum model of a simple solid the definition of free energy is more involved.  We will follow the standard formalism of the continuum mechanics of finite deformation\cite{ogden97,gurtin10}.  Integration is carried in a reference space.  Indicating variables in reference space by a subindex $R$ this integral is expressed as
\begin{equation}
\mathcal{F} = \int_{V_{\mathrm{R}}} \psi_{\mathrm{R}}^{\mathrm{S}} d v_{\mathrm{R}}
\label{intFS}
\end{equation}
$dv_{\mathrm{R}}$ is a volume element in reference space and $V_{\mathrm{R}}$ the corresponding total volume.    Notation is an important technical aspect of finite deformation continuum mechanics.  As in Ref.~\citenum{sprik21c}  we have tried to stay as close as possible  to the notation  of Ref.~\citenum{gurtin10}. 

The infinitesimal volume elements  in Eqs.~\ref{intFL} and \ref{intFS} are related by Jacobian $J$
\begin{equation}
dv = J dv_{\mathrm{R}}
\label{v2vr}
\end{equation}
 $J$ is the determinant of the coordinate transformation between the reference manifold and the space where the deformed body resides.  This transformation is specified by the $3 \times 3$ matrix $\mathbf{F}$ of the derivatives of the Cartesian coordinates $x_i, i=1,2,3$  of a mass point in the deformed body (lower case $x$) with respect to its coordinates $X_i$ in reference space (capital X)
\begin{equation}
 F_{ij} = \frac{\partial x_i }{\partial X_j}
\label{defgrad}
\end{equation}
$\mathbf{F}$ is referred to as the deformation gradient (to unburden notation subindices R have been dropped in Eq.~\ref{defgrad}).  The transformation is invertible 
\begin{equation}
\mathbf{F}^{-1} \mathbf{F} = \boldsymbol{1}
\label{finv}
\end{equation}
$\boldsymbol{1}$ is the unit tensor ($\boldsymbol{1}_{ij} = \delta_{i,j}$).  The determinant $J$ is positive
 \begin{equation}
  J = \det \mathbf{F} > 0
 \label{defJ}
\end{equation}
preserving orientation.

The deformation gradient $\mathbf{F}$ is the central kinematic degree of freedom in non-linear elasticity. A non-linear model is defined by giving an expression for the reference free energy density $\psi_{\mathrm{R}}$ in Eq.~\ref{intFS} in terms of  $\mathbf{F}$
\begin{equation}
\psi_{\mathrm{R}}^{\mathrm{S}} = \hat{\psi}_{\mathrm{R}}^{\mathrm{S}} \left( \mathbf{F}, \rho_{\mathrm{R}} \right) 
\label{psiRS}
\end{equation}
The reference density $\rho_{\mathrm{R}} $ has been added as in explicit state variable in Eq.~\ref{psiRS} as will be needed for the determination of the chemical potential (see section \ref{section:chempot}). $\rho_{\mathrm{R}} $ is related to the density $\rho$ of the deformed body as
\begin{equation}
\rho_{\mathrm{R}} = J \rho
\label{rhor2rho}
\end{equation}
Eq.~\ref{rhor2rho} is a representation of the conservation of the number of particles (Euler equation). The same relation holds between the free energy density in the deformed and reference frame
\begin{equation}
\psi_{\mathrm{R}}^{\mathrm{S}} = J \psi^{\mathrm{S}}
\label{psir2psi}
\end{equation}
However, Eq.~\ref{psir2psi}  is not a conservation law as is Eq.~\ref{rhor2rho}, but a fundamental constitutive assumption  which is referred to as hyperelasticity (see Ref.~\citenum{ogden97}). 

\subsection{From liquids to solids} \label{section:L2S}

A phase boundary in a simple continuum model is a sharp dividing surface marking a constitutive discontinuity.  Each of the free energy densities 
$\psi^{\mathrm{L}}$ for liquid (Eq.~\ref{psiL}) and  $\psi^{\mathrm{S}}_{\mathrm{R}}$ (Eq.~\ref{psiRS}) for the solid phase are therefore stable single well functions.  
Changing a solid in a liquid at this level of approximation is not simply a matter of setting the shear modulus to zero. $\psi_{\mathrm{R}}^{\mathrm{S}}$  represents a "Lagrangian"  energy density in the reference frame while  $\psi^{\mathrm{L}}$  is an "Eulerian" density in the deformed frame. How to bridge constitutive relations in two different manifolds? The approach taken in Ref.~\citenum{sprik21c} is to set up the free energy density of the solid as an Eulerian density just as for the liquid. This density is  then transformed back to the reference frame using Eq.~\ref{psir2psi}  where the stress tensor and chemical potential are computed according to the rules of finite deformation continuum mechanics\cite{gurtin10}.  

The continuum theory as developed here is restricted to mono-atomic systems. The forces in the liquid phase are to a good approximation pair interactions and  short range at the macroscopic length scale.  The free energy density Eq.~\ref{psiL} for such a system is some appropriate well-behaved function  of density $\rho$. 
\begin{equation}
\psi^{\mathrm{L}}= \hat{\psi}^{\mathrm{L}}(\rho) = f^{\mathrm{L}}(\rho)
\label{psihatL}
\end{equation}
The second equality is not a tautology.  After stating the selection of independent mechanical variables in the first equality, the second equality of Eq.~\ref{psihatL} specifies the particular functional dependence defining the constitutive model.  Short range interactions are also present in the solid, similar to the liquid but possibly not identical. To convert the liquid into a solid we add an elastic term of the form used in finite deformation elasticity\cite{ogden97}. 
\begin{equation}
 \psi^{\mathrm{S}} = \hat{\psi}^{\mathrm{S}} \left( \mathbf{F}, \rho_{\mathrm{R}} \right)  =  f^{\mathrm{S}}(\rho) +
 \tfrac{1}{2} G \, \mathrm{tr} \, \mathbf{F}^{\mathrm{T}}\mathbf{F} 
 \label{psihatS}
\end{equation}
$\mathbf{F}^{\mathrm{T}}$ stands for the transpose of the deformation gradient matrix Eq.~\ref{defgrad}. $\mathrm{tr} \mathbf{A}$ is the trace of the matrix $\mathbf{A}$.  The parameter $G>0$ in Eq.~\ref{psihatS} can be regarded as a non-linear version of the Lam\'{e} shear modulus with $f^{\mathrm{S}}(\rho)$ added to account for the bulk modulus (see section \ref{section:stretch}).

The important feature of Eq.~\ref{psihatS} harmonizing liquid and solid is that the argument of $f^{\mathrm{S}}$ is the actual deformed density $\rho$.  The crucial difference marking the solid state is that  $\rho$  is not an  independent mechanical variable.  The independent variables are the deformation gradient $\mathbf{F}$ and the reference density $\rho_{\mathrm{R}}$ as explicitly stated in the first equality of Eq.~\ref{psihatS}.   $\rho$ is an implicit function determined by $\mathbf{F}$ and $\rho_{\mathrm{R}}$  according to Eq.~\ref{rhor2rho}. This construction was chosen to match the isotropic elastic response of the solid to the compressibility in the  liquid. This greatly simplifies the expressions for the gap in chemical potential and grand potential density as we will see later. However, liquids do not support vacancies. Neither does the solid of Eq.~\ref{psihatS} which resembles a liquid in this respect.  Further discussion of this point will be deferred to section \ref{section:limitations}.
 
 For optimal tuning of the constitutive model of the solid to the liquid a further refinement was made. This concerns the second term  in Eq.~\ref{psihatS} accounting for the shear elastic response.  The deformation matrix product   
\begin{equation}
 \mathbf{C} =   \mathbf{F}^{\mathrm{T}}\mathbf{F} 
 \label{Ccg}
 \end{equation}
 is referred to as the right Cauchy-Green deformation tensor. $\mathbf{C}$ is symmetric ($\mathbf{C}^{\mathrm{T}} = \mathbf{C}$) and quantifies non-linear strain\cite{ogden97,gurtin10,gorieli17}
 \begin{equation}
 \mathbf{E}  =  \tfrac{1}{2} \left( \mathbf{C} - \boldsymbol{1} \right) 
 \label{CGstrain}
 \end{equation}
However,  similar to the linear Lam\'{e} model the second term  in Eq.~\ref{psihatS} is not a pure shear but also contributes to the bulk modulus. To separate these two effects  the deformation gradient was factorized  
\begin{equation}
\mathbf{F}  = \mathbf{F}^v  \mathbf{F}^i 
\label{Fvi}
\end{equation} in an  isotropic dilation
\begin{equation}
\mathbf{F}^v  = J^{1/3}  \, \mathbf{I} 
\label{Fv}
\end{equation}
and an isochoric deformation $\mathbf{F}^i$ 
\begin{equation}
\mathbf{F}^i = J^{-1/3} \mathbf{F}
\label{Fi}
\end{equation}
With this definition the determinant of the isochoric factor is unity
\begin{equation}
\det \mathbf{F}^i  =  \left(J^{-1/3}\right)^3 \det \mathbf{F} = 1
\label{detFi}
\end{equation}
The decomposition of the deformation gradient Eq.~\ref{Fvi} is reflected  in a similar factorization of the Cauchy-Green tensor Eq.~\ref{Ccg}. 
\begin{eqnarray}
\mathbf{C} & = & \mathbf{C}^v \mathbf{C}^i
\nonumber \\
\mathbf{C}^v & = & J^{2/3} \boldsymbol{1}
\nonumber \\
\mathbf{C}^i & = & J^{-2/3} \mathbf{C}
\label{Cvi}
\end{eqnarray}
The multiplicative decomposition Eq.~\ref{Fvi} goes back to the continuum mechanics of elastomers where is it used as a procedure to convert an incompressible to a compressible model\cite{ogden97,ogden76,horgan09}.  Multiplicative decomposition is also often applied in the elastic theory of soft matter such as gels\cite{gurtin10,anand11}. 

Adopting this resolution of the deformation gradient  the free energy density Eq.~\ref{psihatS} is written as
\begin{equation}
 \psi^{\mathrm{S}} = \hat{\psi}^{\mathrm{S}}\left(\mathbf{C}^i, J, \rho_{\mathrm{R}} \right)    = 
 f^{\mathrm{S}}\left(\rho \right) + e_{\Lambda}\left(\mathbf{C}^i, J, \rho_{\mathrm{R}}\right)
 \label{liq2sol}
\end{equation}
with $\mathbf{C}^i, J, \rho_{\mathrm{R}}$ as the new independent mechanical variables.  $e_{\Lambda}$ is the isochoric schear elastic energy
\begin{equation}
 e_{\Lambda} \left( \mathbf{C}^i, J, \rho_{\mathrm{R}} \right)  = \tfrac{1}{2} \Lambda(\rho)  \left( \mathrm{tr}  \, \mathbf{C}^i - 3 \right)
 \label{epslamb}
 \end{equation}
 The elastic modulus $\Lambda\left(\rho\right) >0$ is a function of the deformed density $\rho$ in analogy with the short range energy function $f^{\mathrm{S}}(\rho)$.  The free energy of Eq.~\ref{liq2sol} is still an Eulerian density. The  corresponding Lagrangian (reference) density is obtained by multiplying by the determinant $J$ according to Eq.~\ref{psir2psi}. Written out in explicit form we have
 \begin{multline}
 \psi_{\mathrm{R}}^{\mathrm{S}}   =
 \hat{\psi}_{\mathrm{R}}^{\mathrm{S}}   \left( \mathbf{C}^i, J,  \rho_{\mathrm{R}} \right)  = 
   J f^{\mathrm{S}} \left(\rho_{\mathrm{R}}/J \right) 
  \\
   +  \tfrac{1}{2} J  \Lambda \left(\rho_{\mathrm{R}} /J \right)  \left( \mathrm{tr}  \, \mathbf{C}^i- 3 \right)
 \label{psiresf}
 \end{multline}
This is the full hyperelastic energy density used in section \ref{section:soltherm}   for the evaluation of the  chemical potential and stress tensor. The strain energy density Eq.~\ref{psiresf} can be regarded as a variant of a compressible neo-Hookean model\cite{ogden97,horgan09} with the compressibility controlled by the isotropic function $f^{\mathrm{S}}(\rho)$. For later reference the "pull back"  representation of the liquid free energy density Eq.~\ref{psihatL} is added  below 
\begin{equation}
 \hat{\psi}_{\mathrm{R}}^{\mathrm{L}}  = \hat{\psi}_{\mathrm{R}}^{\mathrm{L}}   \left( J,  \rho_{\mathrm{R}} \right) =  J f^{\mathrm{L}} \left(\rho_{\mathrm{R}}/J \right) 
 \label{psirL}
 \end{equation}
 The choice of reference frame for a liquid is arbitrary and should not matter as will be verified in section \ref{section:soltherm}. 
 
 \section{Model thermodynamics}\label{section:soltherm}

\subsection{Chemical potential} \label{section:chempot}
The fundamental thermodynamic derivative relation determining the chemical potential of a solid is
\begin{equation}
\mu = \frac{\partial \hat{\psi}_{\mathrm{R}}\left( \mathbf{F}, \rho_{\mathrm{R}} \right) }{ \partial \rho_{\mathrm{R}}}
\label{mucn}
\end{equation}
Particles are inserted in the reference frame. The elastic deformation $\mathbf{F}$ is kept fixed. The derivation of Eq.~\ref{mucn} is given in the material science literature\cite{cahn78,mullins84,mullins85,sekerka89}.  Taking a somewhat different route the same relation is obtained in the main continuum mechanics text Ref.~\citenum{gurtin10} used as source for our previous work\cite{sprik21c} and again here (see also Ref.~\citenum{gurtin99}). The basic physical reasoning is that solids can distinguish between increasing the number of particles at fixed geometry and the alternative of imposing the same density by changing the shape and volume of a body keeping the particle number the same.  The liquid state which has no memory and cannot detect the difference.  

The derivation of Eq.~\ref{mucn} will not be repeated here. We will only check that the Lagrangian derivative reproduces the familiar thermodynamic relation for liquids determined by the actual (Eulerian) density.   Differentiation of Eq.~\ref{psirL} using the chain rule leads to the result right away
\begin{equation}
\mu^{\mathrm{L}} = \frac{\partial \left(J f^{\mathrm{L}}\left(\rho_{\mathrm{R}}/J \right) \right) }{ \partial \rho_{\mathrm{R}}}
= \frac{d f^{\mathrm{L}}(\rho)}{d \rho}
\label{mul}
\end{equation}
The factors $J$ cancel. The same procedure applies to the first term  in Eq.~\ref{psiresf} defining a short range chemical potential of the solid
\begin{equation}
\bar{\mu}^{\mathrm{S}} = \frac{d f^{\mathrm{S}}(\rho)}{d \rho}
\label{musf}
\end{equation}
The bar has been added because for solids there is a second contribution accounting for the shear elastic energy
\begin{equation}
\mu^{\mathrm{S}} =\bar{\mu}^{\mathrm{S}} +  \kappa_{\Lambda}  e_{\Lambda}
\label{musfe}
\end{equation}
This term is due to the density dependence of shear modulus $\Lambda(\rho)$. The derivative is evaluated as in Eq.~\ref{mul}. The response coefficient $\kappa_{\Lambda}$ given by
\begin{equation}
   \kappa_{\Lambda}  = \frac{1}{\Lambda}\frac{d \Lambda}{d \rho} 
  \label{kappa}
\end{equation}
 $\kappa_{\Lambda}$ can be compared to the electrostriction coefficient in dielectrics. The volume preserving change of shape plays the  role of the electric field.  The  chemical potential is sensitive to shear strain only if the shear modulus varies with density\cite{sprik21a}.  Consistent with the  dielectric parallel, the change in the chemical potential is  proportional to corresponding cost in energy Eq.~\ref{epslamb}.  The coupling of the chemical potential to shear strain is a non-linear effect. 

We are now ready for the grand potential which in section \ref{section:withmu} was found to add a third condition for liquid-solid phase equilbrium. The grand potential density of the solid is as usual defined as
\begin{equation}
\omega^{\mathrm{S}} = \psi^{\mathrm{S}} - \rho \mu^{\mathrm{S}} 
\label{defoms}
\end{equation}
where  $\psi^{\mathrm{S}}$ is the Helmholtz free energy density of Eq.~\ref{liq2sol}.  Both $\psi^{\mathrm{S}}$ and $\rho$ are densities of the deformed solid and therefore $\omega^{\mathrm{S}}$ of Eq.~\ref{defoms} as well.  Substituting Eqs.~\ref{liq2sol} and \ref{musfe} we can write $\omega^{\mathrm{S}}$ as
\begin{equation}
\omega^{\mathrm{S}} = \bar{\omega}^{\mathrm{S}} - \left(  \kappa_{\Lambda} \rho - 1\right)  e_{\Lambda}
\label{omsfe}
\end{equation}
with the short range grand potential density $\bar{\omega}^{\mathrm{S}}$ given by
\begin{equation}
\bar{\omega}^{\mathrm{S}} = f^{\mathrm{S}} - \rho \bar{\mu}^{\mathrm{s}}
\label{baroms}
\end{equation}
The extrapolation of the thermodynamics of liquids to solids seems deceptively easy.  This appealing simplicity is not the result of linear approximations.  Eqs.~\ref{musfe} and \ref{omsfe} are valid for the full non-linear model. This is not a general feature of non-linear thermomechanics  but a consequence  of replacing a solid-like $\rho_{\mathrm{R}}$  dependence  of  $f^{\mathrm{S}}$ and $\Lambda$ by a liquid-like variation with the actual density $\rho$.  It is specific to our constitutive design. This clearly requires further analysis which is deferred to  \ref{section:limitations}.
   
\subsection{Stress tensor for the liquid phase}\label{section:stressL}
The eventual expression for stress will be of comparable simplicity as Eq.~\ref{musfe} for the chemical potential and Eq~\ref{omsfe}. for the grand potential density. The journey is however much more demanding.  The starting point is the stress tensor in the reference frame, the first Piola-Kirchhoff stress tensor
\begin{equation}
\mathbf{T}_{\mathrm{R}} = \frac{\partial \hat{\psi}_{\mathrm{R}}\left( \mathbf{F}, \rho_{\mathrm{R}} \right) }{ \partial \mathbf{F}} 
\label{Trcn}
\end{equation}
The partial derivative Eq.~\ref{Trcn} is the counterpart of Eq.~\ref{mucn}. This time the reference density is held fixed\cite{gurtin10}.  The corresponding stress tensor in the deformed frame  is obtained by Piola transformation 
\begin{equation}
\mathbf{T} = J^{-1}  \mathbf{T}_{\mathrm{R}} \mathbf{F}^{\mathrm{T}}
\label{Trinv} 
\end{equation}
$\mathbf{T}$ is the Cauchy stress tensor which was indicated by $\boldsymbol{\sigma}$ in section \ref{section:Gibbs}.  The change of notation is to stay in line with Ref.~\citenum{gurtin10} where all the mathematics left out here can be found. 

Again, by way of introduction, we first try Eqs.~\ref{Trcn} and \ref{Trinv}  out on the liquid and verify that the detour through reference space ends up with the familiar expression for the pressure determined by the actual density $\rho$.  Inserting the reference liquid free energy Eq.~\ref{psirL} in the derivative Eq.~\ref{Trcn} gives
\begin{equation}
 \frac{ \partial \hat{\psi}_{\mathrm{R}}^{\mathrm{L}} \left(\mathbf{F},\rho_{\mathrm{R}}\right)} { \partial \mathbf{F} }
 =  J  \left(\frac{ d f^{\mathrm{L}}}{ d \rho}  \right) \frac{ \partial \rho}{\partial \mathbf{F}} 
 + f^{\mathrm{L}}\frac{ \partial J }{\partial \mathbf{F}} 
\end{equation}
The second term of the rhs is evaluated using the Jacobi rule for derivatives of determinants
\begin{equation}
 \frac{\partial J}{\partial \mathbf{F}} =  J \mathbf{F}^{-T}
\label{jacobi}
\end{equation}
which also determines the deformation gradient derivative of the density in the first term
\begin{equation}
 \frac{\partial \rho}{\partial \mathbf{F}} = \rho_{\mathrm{R}} \frac{\partial J^{-1}}{\partial \mathbf{F}}
 = - \rho \mathbf{F}^{-T}
\label{drhodF}
\end{equation}
Adding we obtain the Piola stress tensor for the liquid
\begin{equation}
\mathbf{T}_{\mathrm{R}}  = J \left( - \rho \frac{ d f^{\mathrm{L}}}{ d \rho} + f^{\mathrm{L}} \right) \mathbf{F}^{-T}
\label{Trliq}
\end{equation}
Applying the transformation Eq.~\ref{Trinv} yields the Cauchy stress tensor
\begin{equation}
  \mathbf{T}  = - p^{\mathrm{L}}(\rho)  \mathbf{I}
\label{Tlda}
\end{equation}
where
\begin{equation}
  p^{\mathrm{L}}(\rho) = \rho \frac{ d f^{\mathrm{L}}}{ d \rho} - f^{\mathrm{L}} = \rho \mu^{\mathrm{L}} - f^{\mathrm{L}}
  \label{Plda}
\end{equation}
We have recovered the local pressure  of the simple liquid as determined by the free energy density $f^{\mathrm{L}}(\rho)$. 
 
\subsection{Mechanical pressure  for the solid}\label{section:stressS}
The evaluation of the stress tensor for the solid is a more demanding exercise in tensor calculus.  First we split the Cauchy stress tensor in an isotropic and a residual trace free deviatoric part.
\begin{gather}
\mathbf{T} = -p^{\mathrm{S}}  \mathbf{I} + \mathrm{dev} \mathbf{T} 
\nonumber \\[4pt]
 p^{\mathrm{S}} = -  \tfrac{1}{3} \mathrm{tr} \mathbf{T} 
 \nonumber \\[4pt]
 \mathrm{tr} \left(\mathrm{dev} \mathbf{T} \right)  = 0
 \label{deviatoric}
\end{gather}
The minus sign is inserted to be able to interpret $p^{\mathrm{S}}$ in terms of a pressure (as in Eq.~\ref{Tlda} we have suppressed the phase superindex for the stress tensor).  Determination of $p^{\mathrm{S}}$ and $\mathrm{dev} \mathbf{T}$ proceeds in two steps.  First the multiplicative decomposition of deformation Eq.~\ref{Fvi} is implemented in most general form omitting any constitutive detail.  This derivation is taken over in its entirety from Ref.~\citenum{gurtin10}.   Some of the intermediate equations are listed in appendix \ref{cauchy}.  For more detail we refer to the previous publication Ref.~\citenum{sprik21c}.   In the second step the constitutive model of section \ref{section:nlelast} is introduced.   In this section we will focus on the scalar pressure $p^{\mathrm{S}}$.   The deviatoric tensor is the subject of the next section.

The expression obtained by Gurtin et al.~\cite{gurtin10} for  $p^{\mathrm{S}}$ of Eq.~\ref{deviatoric} is written in the notation of section \ref{section:L2S}
\begin{equation}
p^{\mathrm{S}} =  -\frac{\partial \hat{\psi}_{\mathrm{R}}^{\mathrm{S}}\left( \mathbf{C}^i, J, \rho_{\mathrm{R}} \right)}{\partial J}
\label{pJ}
\end{equation}
Mechanical pressure is the response to isotropic change of volume at fixed material density (sometimes referred to as volume strain).  This statement may sound intuitively obvious, the proof is not\cite{gurtin10}.   However,  more than anything else,   this relation persuaded us to stick with the formal non-linear approach deferring linearization to the very last. It is tailor made for our constitutive design.  We can now reiterate the derivation of the liquid pressure in section  \ref{section:stressL}  for each of the two terms in the reference free energy density Eq.~\ref{psiresf}. Starting with the first term we have
\begin{equation}
\frac{\partial}{\partial J} \left[J f^{\mathrm{S}}(\rho_{\mathrm{R}}/J)\right] = f^{\mathrm{S}} - \rho \frac{ d f^{\mathrm{S}}}{d \rho} \equiv
- \bar{p}^{\mathrm{S}}
\end{equation}
On account of Eq.~\ref{musf}  for $\bar{\mu}^{\mathrm{S}}$ the short range pressure can be equally written as
\begin{equation}
\bar{p}^{\mathrm{S}} = \rho \bar{\mu}^{\mathrm{S}} - f^{\mathrm{S}}
\label{barps}
\end{equation}
The  $J$ derivative of the coefficient $\Lambda$ in the second term in Eq.~\ref{psiresf} consists of two similar terms
 \begin{equation}
\frac{\partial}{\partial J} \left[J \Lambda (\rho_{\mathrm{R}}/J)\right]  = \Lambda - \rho \frac{ d \Lambda }{d \rho}
= \Lambda \left( 1 - \rho \kappa_{\Lambda} \right) 
\end{equation}
where $\kappa_{\Lambda}$ is the chemostriction coefficient defined in Eq.~\ref{kappa}.    Adding we find for the mechanical pressure of the solid
\begin{equation}
p^{\mathrm{S}} =  \bar{p}^{\mathrm{S}} + \left( \rho \kappa_{\Lambda} - 1 \right) e_{\Lambda} 
\label{psfe}
\end{equation}
Isochoric deformation  adds a term proportional to the elastic energy $e_{\Lambda}$, not a derivative of $e_{\Lambda}$.  As suggested by the expression Eq.~\ref{musfe} for  the chemical potential,  shear strain is a non-linear effect for the thermodynamics of solids with mobile atoms.  This is also reflected in the mean pressure which is more surprising.

The mechanical pressure in our model is indeed special.  Inserting \ref{barps} and \ref{psfe}  in Eqs.~\ref{omsfe} yields
\begin{equation}
\omega^{\mathrm{S}} = - p^{\mathrm{S}}
\label{gpliq}
\end{equation}
We immediately recognize this identity from the thermodynamics of liquids.  Furthermore, if $p^{\mathrm{S}}$ of Eq.~\ref{psfe} is used to define a molar free enthalpy written in the notation of section \ref{section:Gibbs} as
\begin{equation}
g^{\mathrm{S}} = f_m^{\mathrm{S}} + p^{\mathrm{S}} v_m^{\mathrm{S}} 
\label{fhs}
\end{equation}
we recover the in physical chemistry crucial thermodynamic equivalence of molar free enthalpy and chemical  potential
\begin{equation}
g^{\mathrm{S}} = \mu^{\mathrm{S}}
\label{fhliq}
\end{equation}
with $\mu^{\mathrm{S}}$ given in Eq.~\ref{musfe}. 

Neither Eq.~\ref{gpliq} or Eq.~\ref{fhliq} are definitions but identities bridging the material derivatives Eq.~\ref{mucn}  and Eq.~\ref{pJ}  and thermodynamic potentials of the deformed system. The trace of the Cauchy stress tensor  evidently acts as a quasi-hydrostatic pressure despite the stress tensor being non-hydrostatic. The parallel extends to isothermal variation of density keeping the shape of the solid body as it is. The corresponding first order change of pressure and chemical potential are related as 
\begin{equation}
 d p^{\mathrm{S}} \big|_{\mathbf{C}^i}= \rho d \mu^{\mathrm{S}} \big|_{\mathbf{C}^i}
\label{gibbsduhem}
\end{equation}
where the condition of fixed isochoric deformation has been explicitly indicated. Eq.~\ref{gibbsduhem}  can be compared to the familiar Gibbs-Duhem relation for the thermodynamics of liquids. Increasing the number of particles  at constant volume is equivalent to a decrease of volume at constant particle number. The proof is given in appendix \ref{GdH}.  Recalling the comment at the end of section \ref{section:chempot}, we suspect this fixed shape Gibbs-Duhem reation to be specific to our model for elastic solids,  convenient but  a simplification.

\subsection{Stretching a rectangular prism}\label{section:stretch}

For the evaluation of the deviatoric part (Eq.~\ref{deviatoric})  of the Cauchy stress tensor we abandon the general approach and adopt the special orthorombic geometry of our model system.  Strain  in finite deformation continuum mechanics  is usually specified in terms of stretches obtained from  the eigenvalues of the Cauchy-Green strain tensor (Eq.~\ref{Ccg}).  This is compulsory material in any textbook on applied solid mechanics (Our favourite text at the moment is   Ref.~\citenum{lubarda20}).  The rectangular prism considered here remains oriented along the three orthogonal body axes of the reference frame  and stretch is defined as the ratio
 \begin{equation}
 \lambda_i   = l_i/L_i , \quad i=1,2,3
 \label{stretchi}
\end{equation}
where $l_i$ is the length of the elongated or contracted edge $i$ with $L_i$ the length in the undeformed state. The Cauchy-Green tensor in this geometry is diagonal
\begin{equation}
\mathbf{C} = \begin{pmatrix} \lambda_1^2 & 0 & 0 \\[4pt] 0 & \lambda_2^2 & 0 \\[4pt]  0&  0&  \lambda_3^2 \end{pmatrix}
\label{CGrangle}
\end{equation}
Determinant $J$ of Eq.~\ref{defJ}  is the product of the three stretches
\begin{equation}
   J_{\lambda} = \lambda_1 \lambda_2 \lambda_3 = \sqrt{\det \mathbf{C}}
\end{equation}
An extra suffix $\lambda$ has been appended as a reminder that $J$ is fully determined by the stretches Eq.~\ref{stretchi}.  

The functional form of the elastic energy of an isotropic material is restricted by symmetry\cite{ogden97,gorieli17,lubarda20}. The energy can be expressed is a function of three cubic invariants
\begin{align}
I_1 & = \mathrm{tr} \, \mathbf{C} = \lambda_1^2 + \lambda_2^2 + \lambda_3^2  \nonumber \\[4pt]
I_2  & = \small{\frac{1}{2}}\left[ \left( \mathrm{tr} \, \mathbf{C} \right)^2 -  \mathrm{tr} \, \mathbf{C}^2 \right] =
\lambda_1^2 \lambda_2^2  + \lambda_2^2 \lambda_3^2  + \lambda_3^2 \lambda_1^2 \nonumber \\[4pt]
I_3  & = \det \mathbf{C} = \lambda_1^2 \lambda_2^2  \lambda_3^2 
\end{align}
The non-linear model described by Eq.~\ref{liq2sol} is isotropic and the symmetry rule should apply. This is cleary true for the short range term 
$f^{\mathrm{S}}(\rho)$.  The spatial density $\rho$ is determined by $I_3$ given a reference density $\rho_{\mathrm{R}}$  which is treated as a separate state variable independent of deformation(Eq.~\ref{rhor2rho}).   The elastic energy density Eq.~\ref{epslamb} can be written as
\begin{equation}
e_{\Lambda}   =  \frac{ \Lambda}{2J_\lambda^{2/3} } \,
 \left( \lambda_1^2 + \lambda_2^2 + \lambda_3^2  - 3 \left(\lambda_1 \lambda_2 \lambda_3\right)^{2/3}  \right)
 \label{espstretch} 
\end{equation}
containing  $I_1$ and $I_3$ and a further implicit dependence on $I_3$ because we have allowed $\Lambda$ to vary with $\rho$. This also takes care of the  expression for the mechanical pressure $p^{\mathrm{S}}$  of Eq.~\ref{psfe}.

The detailed expression for the deviatoric part is given in Eq.~\ref{devTCauchyvi} of Appendix \ref{cauchy}.  Evaluation for our model  takes  a bit of effort.  The result we found in Ref.~\citenum{sprik21c} is
\begin{widetext}
\begin{equation}
 \mathrm{dev} \, \mathbf{T} = \frac{\Lambda}{3 J_\lambda^{2/3} }
\begin{pmatrix} 2 \lambda_1^2 - \lambda_2^2 - \lambda_3^2 & 0 & 0 \\
 0 & 2 \lambda_2^2 - \lambda_1^2 - \lambda_3^2 & 0 \\
 0 & 0 & 2 \lambda_3^2  - \lambda_1^2 - \lambda_2^2 \end{pmatrix}
\label{devT}
\end{equation}
\end{widetext}
The trace of the tensor of Eq.~\ref{devT} is clearly zero as required for deviatoric stress.  Cyclic permutation of $\lambda_1, \lambda_{2}, \lambda_{3}$ leads to a  cyclic rearrangement of the components of $ \mathrm{dev} \, \mathbf{T}$ reflecting the cubic symmetry. 

Having derived the thermochemical response  in rigorous non-linear form, the final step is linearizing the stress strain relation Eq.~\ref{devT} for application in section \ref{section:gt}. Linear stress-strain relations are formulated in terms of the infinitesimal strain tensor $\boldsymbol{\epsilon}$ which for our system is of the form $\epsilon_{ij} = \epsilon_i \delta_{ij}$. The linear approximation to the stress Eq.~\ref{devT} is obtained by setting 
\begin{equation}
\lambda_i = 1 + \epsilon_i
\label{lamb2eps}
\end{equation}
and expanding to first order in $\epsilon_i$. The result is
\begin{widetext}
 \begin{equation}
 \mathrm{dev} \, \mathbf{T} = \frac{2 \Lambda}{3}
\begin{pmatrix} 2 \epsilon_1 - \epsilon_2 - \epsilon_3 & 0 & 0 \\
 0 & 2 \epsilon_2 - \epsilon_1 - \epsilon_3 & 0 \\
 0 & 0 & 2 \epsilon_3  - \epsilon_1 - \epsilon_2 \end{pmatrix} 
 \label{deveps}
\end{equation}
\end{widetext}
where we have made the additional approximation that $\Lambda$ is constant ($\kappa_{\Lambda}$ of Eq.~\ref{kappa} is set to zero). We will also need the corresponding small deformation limit of the elastic energy density of Eq.~\ref{epslamb}. Substituting Eq.~\ref{lamb2eps} and leaving out the higher order terms gives  
\begin{equation}
e_{\Lambda} = \frac{ 2 \Lambda}{3}  \left( \epsilon_1^2 + \epsilon_2^2 + \epsilon_3^2 
  - \left( \epsilon_1 \epsilon_2 + \epsilon_1 \epsilon_3 + \epsilon_2 \epsilon_3 \right) \right)
\label{epsquadr}
\end{equation}
The elastic energy induced by isochoric strain  is  a quadratic function of $\epsilon_i$.  The Hessian has a zero eigenvalue (eigenvector $(1,1,1)$) and two degenerate eigenvalues of $ \Lambda$.  The energy Eq.~\ref{epsquadr}  is therefore non-negative. (see note Ref.~\citenum{mistake} and the appendix \ref{lame}).

\section{Melting under non-hydrostatic stress}\label{section:application}

\subsection{Model system setup}\label{section:geometry}
The non-linear elastic model presented in section \ref{section:nlelast} is identical to the model developed in Ref.~\citenum{sprik21c}. The evaluation of the thermodynamic quantities in section \ref{section:soltherm} is also a revision extended with the grand potential density which was not discussed in the previoud publication.  This model will now be applied to a system again with a similar constrained geometry as the example treated in Ref.~\citenum{sprik21c}.  A one-component solid  is enclosed in a rigid box which is open only on one side exposing the  solid to its melt. The solid-liquid interface is a sharp dividing surface.  Interface energies between solid and liquid and with the walls of the container are neglected.  Contact angles are 90 deg  and the interface is assumed to be planar. There are no confinement effects. In this restricted  geometry two of the stretches Eq.~\ref{stretchi} are fixed, say in directions 2 and 3.  Hence $\lambda_2 = \lambda_3 = 1$.  The only mechanical degree of freedom is the ratio $\lambda_1$  between the actual length $l_1$ of the deformed solid sample and the length $L_1$  in the undeformed reference state.  

We will repeat the calculation of Ref.~\citenum{sprik21c}.  We  imagine that the liquid and solid are separated by a piston on top of which we place a weight. The piston is permeable  to ensure chemical equilibrium. Also the rigid walls can be made permeable to allow for exchange of material between the solid inside and the surrounding liquid reservoir outside. The piston can be displaced in the direction perpendicular to its surface. The question is how the system will respond to the applied excess  pressure. This was also investigated in Ref.~\citenum{sprik21c} but what we failed to take into account is that there two modes of displacement for the liquid-solid interface. The solid can be uni-axially compressed or stretched.  This is mechanical deformation controlled by the balance of stress.  Alternatively the surface can migrate. This is an example of the accretion process introduced by Gibbs as explained in section \ref{section:Gibbs}.  

The phase equilibrium conditions obtained in section \ref{section:Gibbs} are the two thermodynamic equalities Eq.~\ref{mugap} and \ref{omegap}  requiring  $\mu^{\mathrm{S}} = \mu^{\mathrm{L}}$  and 
$\omega^{\mathrm{S}} = \omega^{\mathrm{L}}$  complemented by the force balance Eq.~\ref{pgap}.  In  the constrained setup of our model  this only involves the stress component $\sigma_1$.   Subjecting the interface to an external load $p_a$ adds a term to the liquid side of Eq.~\ref{pgap}
\begin{equation}
T_{11} = - p^{\mathrm{L}} - p_a
\label{paload}
\end{equation}
The symbol for the Cauchy stress has been adapted in accordance with the notation used in section \ref{section:soltherm}.

\subsection{Conflicting equilibrium conditions} \label{section:pomegap}

Starting from full hydrostatic phase equilibrium between solid and liquid the load is turned on.  In this initial state $\mu^{\mathrm{S}} = \mu^{\mathrm{L}} = \mu_{\mathrm{H}}$ and $p^{\mathrm{S}} = p^{\mathrm{L}} = p_{\mathrm{H}}$.  We will first consider the case in which the  fluid acts as a reservoir.  The thermodynamic state of the liquid is held constant at $p^{\mathrm{L}} = p_{\mathrm{H}}$ and because the process is strictly isothermal also  $\mu^{\mathrm{L}} = \mu_{\mathrm{H}}$.  This is a severe constraint.  To maintain chemical equilibrium with the fluid the chemical potential in the stressed solid can not deviate from its initial hydrostatic value $\mu^{\mathrm{S}}_{\mathrm{H}} = \mu_{\mathrm{H}}$. Hence 
$ \Delta  \mu^{\mathrm{S}} = \mu^{\mathrm{S}}  -  \mu^{\mathrm{S}}_{\mathrm{H}}  $ must vanish. This implies according to Eq.~\ref{musfe} that
\begin{equation}
  \Delta  \bar{\mu}^{\mathrm{S}} + \kappa_{\Lambda} e_{\Lambda} = 0
  \label{dmubar}
\end{equation}
where $\Delta \bar{\mu}^{\mathrm{S}}$ is the change in the short range atomic chemical potential. $\Delta  e_{\Lambda} = e_{\Lambda}$ because there is no shear elastic energy under hydrostatic pressure.   $e_{\Lambda}$  of the stressed solid is compensated by the  increment $ \Delta \bar{\mu}^{\mathrm{S}}$.   Under isothermal conditions this can only be achieved by adjusting the density $\rho$ of the solid. Expanding Eq.~\ref{dmubar} to  first order in $\Delta \rho^{\mathrm{S}} = \rho^{\mathrm{S}} - \rho_{\mathrm{H}}$ ignoring the density dependence of the response coefficient $\kappa_{\Lambda}$ Eq.~\ref{dmubar} becomes
\begin{equation}
 \Delta \rho^{\mathrm{S}} = -  \rho_{\mathrm{H}}^2 \kappa_{\rho} \kappa_{\Lambda} e_{\Lambda} 
 \label{drhoload}
\end{equation}  
 The coefficient $\kappa_{\rho}$ defined as
\begin{equation}
\frac{1}{\kappa_{\rho}} = \rho \frac{d \bar{p}^{\mathrm{S}}}{d \rho} = \rho^2 \frac{d \bar{\mu}^{\mathrm{S}}}{d \rho}
\label{kapparho}
\end{equation} 
is the  isothermal compressibility. 

The chemical equilibrium condition Eq.~\ref{drhoload} is highly  restrictive for systems without the freedom of chemostriction ($\kappa_\Lambda =0$).  The chemical potentials of the solid and liquid can only match  if the density $\rho^{\mathrm{S}}$ in the non-hydrostatic state has not changed from its value $\rho_{\mathrm{H}}$ in the initial hydrostatic state.  For finite $\kappa_{\Lambda} $ the solid responds to non-hydrostatic stress by a change of density proportional to the elastic energy.  However, there are further requirements for equilibrium.  The system must still satisfy the condition of equal  grand potential density.   Because of the connection between grand potential density and mechanical pressure (Eq.~\ref{gpliq})  this translates into a mechanical pressure balance
\begin{equation}
 \Delta \bar{p}^{\mathrm{S}} + \left(  \rho^{\mathrm{S}} \kappa_{\Lambda} - 1 \right) e_{\Lambda} = 0
\label{dpbar}
\end{equation}
Eq.~\ref{dpbar} is an independent thermodynamic constraint in addition to  Eq.~\ref{dmubar}.   

The  $\kappa_\Lambda =0$  limit of Eq.~\ref{dpbar}  is again  straightforward and will be investigated first.  Eliminating the term proportional to $\kappa_{\Lambda}$  from  Eq.~\ref{dpbar} we have
\begin{equation}
\Delta \bar{p}^{\mathrm{S}} =  e_{\Lambda} 
\label{dpbar0}
\end{equation}
However, recalling Eq.~\ref{drhoload},  changes in density are only permitted under chemical equilibrium for finite chemostriction.   Without chemostriction the short range pressure $\bar{p}^{\mathrm{S}}$ is locked to the  hydrostatic pressure $p^{\mathrm{L}} = p_{\mathrm{H}}$ of the liquid reservoir.  $\Delta \bar{p}^{\mathrm{S}} = 0$ and therefore also $e_{\Lambda}$.  Non-hydrostatic pressure is incompatible with thermochemical equilibrium without coupling between density and shear stress.  As will be shown in appendix \ref{GdH}  the incompatibility between chemical and thermodynamic pressure balance cannot be resolved by relaxing the $\kappa_{\Lambda} = 0$ constraint.  This is what went wrong in our previous study Ref.~\citenum{sprik21c}.  Enforcing chemical equilibrium, but unaware of the violation of the stability condition for  interface migration, we went ahead and  solved the force balance equation Eq~\ref{paload} and made the unjustified  conclusion that finite strain induced by the load $p_a$ is  allowed under chemical equilibrium.

\subsection{Gibbs-Thomson pressure correction} \label{section:gt}
The solid in the experiment of section \ref{section:pomegap} was initially in hydrostatic equilibrium with its liquid phase before driven away from coexistence by the application of non-hydrostatic pressure.  The pressure  $p^{\mathrm{L}}$ of the liquid was held fixed at the hydrostatic value $p_{\mathrm{H}}$.   Can phase equilibrium be restored by adjusting the thermodynamic state of the liquid?  These corrections are explicitly dependent on the stress tensor in the solid body and are known as Gibbs-Thomson (GT) equations\cite{sekerka04,voorhees04}.  Intended for temperature control, similar corrections are also used  for tuning the hydrostatic pressure of the liquid at constant temperature\cite{mishin10,speck24}.  These pressure offsets can be compared  to the Kelvin correction to the vapour pressure  at  phase equilibrium with finite size droplets of the liquid condensate. The shear elastic energy  plays the same role as the surface tension at the liquid-vapour interface with the important difference is that GT corrections for nonhydrostatic stress remain finite for planar interfaces.  

GT corrections are obtained applying  perturbation methods to the Gibbs phase equilibrium condition Eq.~\ref{muparadox}.  As in the original derivation by Gibbs chemical equilibrium is usually not explicitly taken into account\cite{mishin10,speck24,nozieres93}.  As we have seen,  combining particle exchange with the Gibbs accretion leads to two zero gap conditions, one for the grand potential density (Eq.~\ref{omegap}) and one for the chemical potential (Eq.~\ref{mugap}).  This raises the question whether the strengthening of the phase equilibrium conditions  will alter the GT pressure corrections.   Similar to liquids the grand potential for a solid under non-hydrostatic pressure is still equal  to minus the mechanical pressure $p^{\mathrm{S}}$.  We reiterate that this property is specific to our nonlinear continuum model.  To maintain equilibrium the change $\Delta p^{\mathrm{S}}$ must match the change  $\Delta p^{\mathrm{L}} = p^{\mathrm{L}}  - p_{\mathrm{H}}$ of the liquid.  In stead of zero on the rhs of Eq.~\ref{dpbar} we now have 
\begin{equation}
\Delta \bar{p}^{\mathrm{S}} + \left( \kappa_{\Lambda} \rho^{\mathrm{S}} -1 \right) e_{\Lambda} = \Delta p^{\mathrm{L}} 
\label{dpbargt}
\end{equation}
Also the chemical potentials must remain in balance and Eq.~\ref{dmubar} is modified to
\begin{equation}
\Delta \bar{\mu}^{\mathrm{S}} + \kappa_{\Lambda} e_{\Lambda} = \Delta \mu^{\mathrm{L}}
\label{dmubargt}
\end{equation}
where $\Delta \mu^{\mathrm{L}} = \mu^{\mathrm{L}} - \mu_{\mathrm{H}}$.  As shown in appendix \ref{section:gtmath}  Eqs.~\ref{dpbargt} and \ref{dmubargt} the GT pressure correction is proportional to the shear strain energy $e_{\Lambda}$ in the solid
\begin{equation}
 \Delta p^{\mathrm{L}} 
 = \left( 1 - \kappa_{\Lambda} \Delta \rho^{\mathrm{S}}  \right) 
  \frac{\rho_{\mathrm{H}}^{\mathrm{L}} e_{\Lambda}}{\rho_{\mathrm{H}}^{\mathrm{S}} -  \rho_{\mathrm{H}}^{\mathrm{L}} } 
 \label{gteps}
\end{equation}
Finite chemostriction ($\kappa_{\Lambda} \neq 0$)  adds a prefactor depending on the change in solid density $\Delta \rho^{\mathrm{S}} = \rho^{\mathrm{S}} -  \rho_{\mathrm{H}} ^{\mathrm{S}}$ induced by the coupling between the shear and volume strain.    

 As mentioned, the GT pressure correction is derived in Refs.~\citenum{mishin10} and \citenum{speck24} directly 
from the fundamental  Gibbs accretion equilibrium condition Eq.~\ref{mulaccr} converted to a  density representation
\begin{equation}
\frac{\psi^{\mathrm{S}}}{\rho^{\mathrm{S}}} + \frac{p^{\mathrm{L}}}{\rho^{\mathrm{S}}}  = \mu^{\mathrm{L}}
\label{mulaccrv}
\end{equation}
Carried out for the specific setup of our constititive model this procedure leads to 
\begin{equation}
  \Delta p^{\mathrm{L}} 
 =   \frac{\rho_{\mathrm{H}}^{\mathrm{L}} e_{\Lambda}}{\rho_{\mathrm{H}}^{\mathrm{S}} -  \rho_{\mathrm{H}}^{\mathrm{L}} } 
 \label{gteps0}
\end{equation}
The proof is given in appendix \ref{section:gtmath}.  The GT pressure of Eq.~\ref{gteps0} is the zero in the chemostriction limit of Eq.~\ref{gteps}. Evidentty chemical equilibrium is only a concern when the density is physically coupled to shear strain.

The expression Eq.~\ref{gteps} for the GT correction is generally valid. Applied surface forces (tractions) and boundary conditions are implicit in the actually value of the strain energy  $e_{\Lambda}$ as determined by the Cauchy force balance. For our simple constrained geometry this is Eq.~\ref{paload}. 
Taking the increment in the liquid pressure into account the force balance equation Eq.~\ref{paload} is extended to
\begin{equation}
T_{11} =-p_{\mathrm{H}} - \Delta p^{\mathrm{L}} - p_a
\label{gtpaload}
\end{equation}
Eq.~\ref{deviatoric} gives for the normal stress on the solid side of the interface 
\begin{equation}
T_{11} = - p^{\mathrm{S}} + \left(\mathrm{dev} \mathbf{T}\right)_{11}
\end{equation}
On account of the thermodynamic pressure balance $p^{\mathrm{S}}$ can be replaced by $p^{\mathrm{L}} = p_{\mathrm{H}} + \Delta p^{\mathrm{L}}$ and we obtain
\begin{equation}
T_{11} = - p_{\mathrm{H}} - \Delta p^{\mathrm{L}}+ \left(\mathrm{dev} \mathbf{T}\right)_{11}
\end{equation}
Substituting in Eq.~\ref{gtpaload} yields
\begin{equation}
\left(\mathrm{dev} \mathbf{T}\right)_{11} = - p_a
\end{equation}
Now resorting to  linearization, the  $(11)$ component of the deviatoric Cauchy stress tensor is computed by setting  $\epsilon_2 = \epsilon_3 = 0 $ in Eq.~\ref{deveps}. This gives
\begin{equation}
 \left(\mathrm{dev} \mathbf{T}\right)_{11} = \frac{4 \Lambda}{3} \epsilon_1
\label{Tlin}
\end{equation}
Inserting in the force balance  Eq.~\ref{paload} we find
\begin{equation}
\epsilon_1 = -  \frac{3 p_a}{4 \Lambda}
\label{pastretch}
\end{equation}
The thermodynamic pressure equilibrium has cancelled out the dependence on the pressure in the liquid. 
The strain Eq.~\ref{pastretch}  is identical to what we found in Ref.~\citenum{sprik21c}.  
 
 What remains is determining elastic energy generated by the strain Eq.~\ref{pastretch} applying Eq.~\ref{epsquadr}
 \begin{equation}
e_{\Lambda} = \frac{ 2 \Lambda}{3}  \epsilon_1^2 =  \frac{3 p_a^2}{8 \Lambda}
\label{epstrecth}
\end{equation}
and then substituting in the equation for the GT pressure Eq.~\ref{gteps}. Ignoring chemostriction we find
\begin{equation}
 \Delta p^{\mathrm{L}} =  \frac{ 3 \rho_{\mathrm{H}}^{\mathrm{L}} p_a^2 }{8 \Lambda \left( \rho_{\mathrm{H}}^{\mathrm{S}} -  \rho_{\mathrm{H}}^{\mathrm{L}} \right) }  
 \label{gtpaL}
\end{equation}
The GT pressure for our model system is inversely proportional to the shear modulus. Alternatively we can rewrite Eq.~\ref{gtpaL} in terms of 
Young modules $E$ and Poisson ratio $\nu$ (see Appendix \ref{lame})
\begin{equation}
\Delta p^{\mathrm{L}} =  \frac{ 3 \left(1 + \nu \right) \rho_{\mathrm{H}}^{\mathrm{L}} p_a^2 }{4 E \left( \rho_{\mathrm{H}}^{\mathrm{S}} -  \rho_{\mathrm{H}}^{\mathrm{L}} \right) }  
\label{gtpaEvu}
\end{equation}
In this form our result can be directly compared to the expression given by  Nozi\`{e}res\cite{nozieres93,nozieres95}
\begin{equation}
\Delta p^{\mathrm{L}} =  \frac{ \left(1 - \nu^2 \right) \rho_{\mathrm{H}}^{\mathrm{L}} p_a^2 }{ E \left( \rho_{\mathrm{H}}^{\mathrm{S}} -  \rho_{\mathrm{H}}^{\mathrm{L}} \right) }  
\label{gtpanoz}
\end{equation}
These two equations for the GT pressure are not the same. The variation with the Poisson ratio is qualitatively different, increasing for Eq.~\ref{gtpaEvu} and decreasing for Eq.~\ref{gtpanoz}. The reason for this difference is that the excess stress in Ref.~\citenum{nozieres93,nozieres95} is applied parallel to the solid/liquid interface instead of directly loading the interface itself as in our system. 

 Eq.~\ref{gtpanoz}  was obtained  by Nozi\`{e}res assuming isotropic elasticity.   General expressions also valid for anistropic elastic response are given in Refs.~\citenum{mishin10} and \citenum{speck24}.   Eq.~\ref{gtpanoz} is obtained by substituting the compliances Eq.~\ref{compliance}.  The GT theory was verified against extensive molecular dynamics simulations in Refs.~\citenum{mishin10} and \citenum{speck24}.  The model consists of a system of atoms in a rectangular periodic box with a pair of opposite solid liquid interfaces.  Applying bi-axial stress, the stability of these interfaces is investigated.   The change in equilibrium state is compared to the  GT relations as presented in these publications.  The authors of Ref.~\citenum{mishin10} and \citenum{speck24} find that their numerical results are in good agreement with the theory.  The issue of a chemical potential for the solid phase is not discussed but it can be assumed that the molecular dynamics systems are in chemical equilibrium.  This suggest that the chemostriction correction in Eq.~\ref{gteps}  is a very minor effect.  Non-linear corrections are indeed negligible under the conditions the simulations were carried out. 

\section{Wider context and limitations}\label{section:discussion}
\subsection{Configurational mechanics} \label{section:conforce}

Before discussing the limitations of our constitutive model it may be instructive to return to the solid-liquid  phase equilibrium  thermodynamics underlying the results presented in section \ref{section:application}. This concerns in particular  the  distinction between two types of motion of a dividing surface, deformation and migration. The fundamental questions raised by treating interface migration as a separate degree of freedom have been the focus of research in theoretical continuum mechanics for most of  the second half of the last century  leading to the introduction of configurational forces.  A selection of often quoted reviews\cite{gurtin95,gurtin99} and books\cite{maugin93,gurtin00,kienzler00,maugin11} is included in the references.  Making contact with this profound development may hopefully help clarifying the results for our model system.  However, the considerations in this section are tentative  and at this stage must be regarded as no more than a philosophical afterthought. 

The origins of configurational mechanics are credited to Eshelby.  As defined in his pioneering papers,  configurational stress is due to rearrangement of a material  structural feature such as a defect or an internal interface\cite{eshelby56,eshelby75}.  In the strong  formulation  of the theory configurational stress is conceptually different from standard elastic stress  with separate balance laws.  This statement  has led to some controversy\cite{podioguidugli02,fried06}.  Another issue is how to merge configurational forces with thermodynamic forces,  in particular chemical potentials.  Here we will follow the interpretation of Gurtin and Fried  who link configurational forces to the accretion force introduced by Gibbs.\cite{gurtin95,gurtin99,gurtin00}. 

Returning to our model system, consider the question how to quantify the distinction between deformation and interface migration. This seems possible using the freedom in the definition  of stretch of Eq.~\ref{stretchi}.   Increasing $l_1$ at fixed $L_1$ corresponds to the usual elastic stretching.  Alternatively, the same strain (the ratio $l_1/L_1$) can be generated by reducing $L_1$ at fixed $l_1$.  This would amount to modifying the reference configuration while preserving the spatial configuration specified in our system by $l_1$.   As a mechanical process this could be viewed as reverse motion.  However, from a thermodynamic perspective, moving an interface back in reference space is not the same as moving it forward by deformation. The reference space  process involves transfer of material across the interface. This is clearly illustrated by the Gibbs accretion procedure of section \ref{section:nomu}.  Elastic stretching, on the contrary, conserves the mass on either side of the surface. This suggest that a force acting on  $L_1$  at fixed $l_1$ can be regarded as a configurational force.  This motion equally strains the solid,  but the energies involved are partly thermodynamic and will not be the same, hence the difference in phase equilibrium condition.

  \subsection{Defects and surface energies missing} \label{section:limitations}
There are obvious limitations to the constitutive model of the solid.  Liquids do not support vacancies, solids do. This is a second fundamental distinction between solids and liquids in addition to shear elasticity\cite{cahn85,mullins85,larche96,voorhees04,martinpc72,fleming76,fuchs15}.  The energy of a lattice with vacancies differs from the energy of a perfect lattice with the identical  average density free of vacancies\cite{frenkel07}.  This is a characteristic property of a crystal independent of shear strain.  However, this distinction is lost in our liquid-like model. The average density is the only state variable under hydrostatic stress.  One could take the strict point of view that the model is not describing a crystal,  but some kind of elastic fluid.\cite{cahn85}.  Restoring the formation of vacancies will therefore require treating the deformation gradient determinant $J$ and reference density $\rho_{\mathrm{R}}$ as genuine independent degrees of freedom instead of combining  them in the deformed density $\rho= \rho_{\mathrm{R}}/J$. This will bring substantial changes to the solid thermodynamics.  In particular,  the for a solid suspicious equivalence Eq.~\ref{gpliq} between grandpotential density and mechanical pressure no longer applies.
 
To clarify this point with a simple example,  let us explore briefly what happens if the reference energy density Eq.~\ref{psiresf} is extended with a term $f_{\mathrm{R}}(\rho_{\mathrm{R}})$ varying only with  material density $\rho_{\mathrm{R}}$
 \begin{multline}
 \psi_{\mathrm{R}}^{\mathrm{S}}   = 
   J f^{\mathrm{S}} \left(\rho_{\mathrm{R}}/J \right)  + f_{\mathrm{R}}\left(\rho_{\mathrm{R}}\right)
  \\
   +  \tfrac{1}{2} J  \Lambda \left(\rho_{\mathrm{R}} /J \right)  \left( \mathrm{tr}  \, \mathbf{C}^i- 3 \right)
 \label{psiresfv}
 \end{multline}
 The extra term is picked up by the reference density derivative Eq.~\ref{mucn} determining the chemical potential. This amounts to adding a term
 \begin{equation}
 \mu_{\mathrm{R}}= \frac{ d f_{\mathrm{R}} \left( \rho_{\mathrm{R}} \right)}{d \rho_{\mathrm{R}}}
 \end{equation}
 to the expression Eq.~\ref{musfe} for the chemical potential
 \begin{equation}
\mu^{\mathrm{S}} =\bar{\mu}^{\mathrm{S}} +  \kappa_{\Lambda}  e_{\Lambda} + \mu_{\mathrm{R}}
\label{musfev}
\end{equation}
$\mu_{\mathrm{R}}$ can be regarded as an intrinsic contribution to the  chemical potential due to vacancies and other defects.  Adjusting the chemical equilibrium Eq.~\ref{mugap}  we have
\begin{equation}
\mu^{\mathrm{L}}- \mu^{\mathrm{S}} = \mu_{\mathrm{R}}
\label{murgap}
\end{equation}
A gap between chemical potentials of solid and liquid forbidden in the old model is now allowed  provided it can be balanced by the intrinsic  chemical potential $\mu_{\mathrm{R}}$ of the solid.  Note that Eq.~\ref{murgap} is equally valid for interfaces under hydrostatic pressure. 



In contrast,   the modification Eq.~\ref{psiresfv} has no effect on the mechanical pressure.  Energy densities depending exclusively on reference density are  skipped by the $J$ derivative in Eq.~\ref{pJ}.  The mechanical pressure is still given by Eq.~\ref{psfe}.  This is not true for the grand potential density, which is modified to 
\begin{equation}
\omega^{\mathrm{S}} = - p^{\mathrm{S}} - \frac{p_{\mathrm{R}}}{J}
\label{omsfev}
 \end{equation}
 with $p_{\mathrm{R}}$ given by 
\begin{equation}
p_{\mathrm{R}} = \mu_{\mathrm{R}} \rho_{\mathrm{R}} - f_{\mathrm{R}}
\end{equation}
$p_{\mathrm{R}}$ can be interpreted as an effective material pressure generated by $f_{\mathrm{R}}$.   As a result the "pseudoliquid" equivalence Eq.~\ref{gpliq} between grandpotential density and mechanical pressure is lifted.   The scaling factor in  Eq.~\ref{omsfev} is the inverse volume strain $1/J = V_{\mathrm{R}}/V$ enhancing the effect  of defects in compressed solids.  

Substituting Eq.~\ref{omsfev} in the Gibbs phase equilibrium condition $\omega^{\mathrm{L}} = \omega^{\mathrm{S}}$ opens up a finite gap between the thermodynamic pressures as determined by the non-linear elastic model.  Again this also applies to a coexisting liquid and solid phases under nonhydrostatic pressure.  In absence of shear stress the equilibrium equation for thermodynamic pressure is simplified to a thermodynamic jump condition for hydrostatic pressure
\begin{equation}
p^{\mathrm{L}} - \bar{p}^{\mathrm{S}} = \frac{p_{\mathrm{R}}}{J}
\label{phrgap}
\end{equation}
 We close this appealing but distinctly speculative argument with the warning  that the introduction of the material potential $f_{\mathrm{R}}$ in Eq.~\ref{psiresfv} is merely meant as an indication of the special character of a continuum mechanics missing such a contibution.   To be considered as a realistic option more justification is needed.   For the moment, the physical picture is missing.  What to expect for the magnitude of $\mu_{\mathrm{R}}$ and $p_{\mathrm{R}}$? How does this scheme compare to the network solid theory of Larch\'{e} and Cahn\cite{cahn73,cahn78} and the related non-equilibrium thermodynamics based  approach of Gurtin and Fried\cite{gurtin10,gurtin99}?   Finally, of course,  there is the crucial  question about experimental evidence.  This will be  left as the subject of future research.

A further obvious limitation of our model is that  surface and interface energies have been omitted.  This should not have an effect on the planar interfaces considered here. Only curvature can couple surface tension and surface stress to bulk stress. The well known example is the Young-Laplace law. The condition is however that the surface excess quantities are uniform. Bulk stress is in fact sensitive to modulation of surface stress of planar surfaces as shown by Gurtin and Murdoch\cite{gurtin75,gurtin78}.  For soft solids this can give rise to complicated elasto-capillary effects\cite{style17}.  For the solid-melt interfaces stress driven corrugation could play this role \cite{voorhees04,grinfeld92,grinfeld93,nozieres93,nozieres95,srolovitz89,kassner01}. 
 
 With an upgrade to surface excess energies the model  would also be ready for what is likely to be a more topical application,  theory and simulation of the micro crystals of nucleation theory and solids in pores.  Non-hydrostatic stress in confined solids is well documented and can take surprisingly large values\cite{gubbins11,gubbins14}. The thermodynamic formalism presented in this paper could therefore be relevant for the study of capillary freezing, provided the pores sizes are not completely out of reach of macroscopic modelling.  There is no consensus on the role of non-hydrostatic bulk stress for crystal nucleation.  However,  surface stress and surface tension in the crystalline state can be substantially different (even of opposite sign)\cite{frenkel04}. It has been shown that this difference gives rise to deviations between pressure and grand potential density even if the pressure tensor inside the nucleus is isotropic\cite{mullins84,sekerka89,voorhees04}. As a result the grand potential density gap again must be added to the equilibrium conditions just as for bulk solids under non-hydrostatic stress as indicated by careful simulation work\cite{frenkel04,vega20,vega22}. 

 \section{Recapitulation and conclusion} \label{section:conclusion}

Surfaces and interfaces are all important in the macroscopic physics of solids.  This is why the elementary forces in solid mechanics are  contact forces (tractions) acting at the surface of a body as postulated by Cauchy.  Gibbs generalized Cauchy's tractions to interface forces  realizing that the evolution of internal dividing surfaces is driven by a separate set of accretion forces.  This was subsequently formalized in the modern  theory of configurational  mechanics as briefly indicated in section \ref{section:conforce}.  Chemical potentials were,  initially,  less of a consideration.  However,  changes in composition are crucial in material science,  in particular metallurgy.  The next step was therefore the introduction of chemical potentials.  Leading names in this development are Herring,  Cahn,  Mullins,  Sekerka and coworkers.  Their work unified the continuum mechanics of solids with  the ``Gibbsian'' thermodynamics of phases.   The  Gibbs accretion equilibrium  was reformulated in terms of a relation between  the grand potential energy densities of the two phases which acts along side chemical and mechanical equilibrium conditions\cite{sekerka89,voorhees04}.  This third equilibrium condition makes a difference for solids under non-hydrostatic stress.

The motivation for the study of  Ref.~\citenum{sprik21c} was to investigate what would happen to solid-liquid coexistence if the solid is under non-hydrostatic stress. This question has been, and continues to be,  of prime interest in material science.  However in Ref.~\citenum{sprik21c}  the problem was approached in the regular framework of physical chemistry (the academic background of the author).   Focused on liquids,  the intricacies of the  thermodynamics of solids are not always fully appreciated in physical chemistry.   This biased view dominates, for example,  the molecular dynamics study of crystal nucleation\cite{vega20,vega22,michaelides16,ceriotti17}(exceptions are Refs.~\citenum{frenkel04} and \citenum{haertel12}).  Accordingly,  equalizing chemical potentials,  we set up the balance of stress at the interface using a non-linear elastic  model selected for this purpose.   Technically  the model can be classified as an open compressible neo-Hookean material\cite{ogden97}.  The stress tensor was worked out following the established rules of finite deformation continuum mechanics. In addition,  also the chemical potential was determined applying methods borrowed from the continuum theory of diffusion in solids\cite{gurtin10}.    This scheme was applied to a solid sample in a container as described  in section \ref{section:geometry}.  It was found that the solid in phase equilibrium with its melt in this constrained geometry  can be stable under non-hydrostatic stress.  

However,  the condition of uniform  grand potential density makes a difference for this system but was overlooked.  With the chemical potential in hand we now have derived the required expression for the grand potential density.   It was found that, similar to a liquid, the grand potential density is still equal to minus the mean pressure defined as minus a third of the trace of the stress tensor. This relation persists  even if the pressure is non-hydrostatic which turned out to be quite convenient for enforcing the zero grand potential density gap condition for phase equilibrium. Using exactly the same continuum model we now come to the opposite conclusion.  The interface is not stable.   Stability can however be restored by adjusting the pressure of the liquid.  This was shown by deriving an explicit expression for the Gibbs-Thomson pressure correction for the isotropic non-linear model which is the basis for this study.  This is a key result of the present followup  paper. 

From a more fundamental perspective,  we feel that  the continuum model presented here could be helpful for the microscopic understanding of the chemical thermodynamics of non-hydrostatic stress.  Microscopic theory and continuum mechanics approach chemomechanical coupling from opposite ends.  Stress and strain are the basic primitive variables in the continuum mechanics with an unambiguous mathematical foundation.  Assigning a chemical potential to a continuum solid on the other hand,  is a non-trivial problem raising profound questions about the nature of the solid state\cite{cahn85}.  In microscopic theory it is the other way around.   While the chemical potential is a fundamental concept in statistical mechanics\cite{evans79,rowlinson02},  the microscopic definition of stress remains the subject of debate (even of controversy)\cite{kirkwood50,murdoch85,mistura85,rowlinson02,debenedetti12,dicarlo21,gubbins23}.   As a consequence the question of computation of the stress tensor in molecular simulation has also not been settled \cite{tildesley16,ersmith22,dicarlo23}.  

The present entirely macroscopic study has little to contribute to these questions other than reminding us of the connection between pressure and the grand potential density\cite{percus93}.  In uniform liquids the grand potential density is equal to minus the hydrostatic pressure.  We have shown that this relation can be generalized to uniform solids under non-hydrostatic stress,  at least for the special non-linear elastic model considered in this paper.   Moreover,  possible mechanisms for the breakdown of this relation,  as discussed in  section \ref{section:limitations}),  are likely to be equally instructive. The identification of grand potential density gradients as driving force for interface migration adds a further aspect to these problems which might be worth considering in a  microscopic framework. The summary of all of this is perhaps that, configurational forces, as very briefly introduced in section \ref{section:conforce}, deserve more attention in statistical mechanics.

\section*{Acknowledgement}
Daan Frenkel is acknowledged for numerous discussions about stress, both confusing and helpful.

\section*{Disclosure statement}
No potential conflict of interest was reported by the author.










\appendix

\section{Stress tensor, more technical detail.}\label{cauchy}
As indicated in section \ref{section:stressL}  the primary stress tensor in finite deformation continuum mechanics is the Piola stress tensor Eq.~\ref{Trcn}. 
The expression derived in  Ref.~\citenum{gurtin10} is  
\begin{multline}
\mathbf{T}_{\mathrm{R}}  = - J p  \mathbf{F}^{-\mathrm{T}}
 \\ 
+  J^{-1/3} \left[ \mathbf{F}^i  \boldsymbol{\Pi}^i   - \tfrac{1}{3} \left( \mathbf{C}^i :  \boldsymbol{\Pi}^i  \right)\mathbf{F}^{i-\mathrm{T}} \right]  
\label{Trvi}
\end{multline}
The double dot tensor product $\mathbf{A} : \mathbf{B}$ stands for the contraction $\sum_{ij} A_{ij} B_{ij}$. Eq.~\ref{Trvi} is a generic identity valid for the multiplicative decomposition Eq.~\ref{Fvi}.   $\mathbf{C}^i$ is the isochoric Cauchy-Green tensor of Eq.~\ref{Cvi}. The factor $p$ is the mechanical pressure of Eq.~\ref{pJ}.  $\mathbf{C}^i$  and $\rho_{\mathrm{R}}$ are kept constant in Eq.~\ref{pJ}. We also encountered already the $\rho_{\mathrm{R}}$ derivative at fixed $\mathbf{C}^i$ and $J$. This  is the chemical potential of Eq.~\ref{mucn}. The tensor $\boldsymbol{\Pi}^i$ is the third partial derivative
\begin{equation}
\boldsymbol{\Pi}^i = 2 \frac{\partial \hat{\psi}_{\mathrm{R}}^{\mathrm{S}}\left( \mathbf{C}^i, J, \rho_{\mathrm{R}}\right)}
{\partial \mathbf{C}^{i} } 
\label{pii}
\end{equation}
which quantifies the response to isochoric deformation. 

Next the Piola stress tensor Eq.~\ref{Trvi}  is transformed to an expression for the Cauchy stress tensor $\mathbf{T}$ applying Eq.~\ref{Trinv}. 
 This right away converts the first term in Eq.~\ref{Trvi} to the isotropic component in Eq.~\ref{deviatoric}.  The second term is found by combining
 Eq.~\ref{Trvi} with Eq.~\ref{Fi}. 
 \begin{equation}
 \mathbf{T} = J^{-2/3} \mathbf{T}_{\mathrm{R}} \mathbf{F}^{i\mathrm{T}}
 \end{equation} 
 Applying the reference to deformed frame transformation in this form we obtain for the deviatoric component of the Cauchy stress tensor
\begin{equation}
\mathrm{dev} \mathbf{T}  =  J^{-1} \left[ \mathbf{F}^i  \boldsymbol{\Pi}^i \mathbf{F}^{i \mathrm{T}}
   - \tfrac{1}{3} \left( \mathbf{C}^i :  \boldsymbol{\Pi}^i  \right)\boldsymbol{1} \right]
 \label{devTCauchyvi}
 \end{equation}
 

\section{Including chemostriction} \label{GdH}
Some of the key conclusions in the main text are based on equations which are valid only for  the chemostriction coefficient $\kappa_{\Lambda}$ set to zero. The example is thermodynamic force balance Eq.~\ref{dpbar0}. Others, such as  Eq.~\ref{gibbsduhem},  were presented without proof.  In this appendix we revisit these equations giving a derivation for finite $\kappa_{\Lambda}$. We start with the fixed shape Gibbs-Duhem relation Eq.~\ref{gibbsduhem}.
Evaluation of the differential of Eq.~\ref{psfe} for the pressure gives
\begin{equation}
d p^{\mathrm{S}} = d \bar{p}^{\mathrm{S}} + \kappa_{\Lambda} e_{\Lambda} d \rho^{\mathrm{S}} + \left( \rho^{\mathrm{S}} \kappa_{\Lambda}  - 1 \right) d  e_{\Lambda} 
\label{dps1}
\end{equation}
where the higher order effect of a possible density dependence of the $\kappa_{\Lambda}$ has been ignored.  To stay in line with the more elaborate notation of section \ref{section:gt} the density of the solid is now also explicitly labeled with a superindex S. The first term of Eq.~\ref{dps1}  is the differential of the short range atomic pressure of Eq.~\ref{barps} which satisfies the usual Gibbs-Duhem relation $d \bar{p} = \rho  d \bar{\mu}$ for liquids. Substituting we can write
\begin{equation}
d p^{\mathrm{S}} = \rho^{\mathrm{S}} \left( d\bar{\mu}^{\mathrm{S}} + \kappa_{\Lambda} de_{\Lambda} \right)  + \kappa_{\Lambda} e_{\Lambda} d \rho^{\mathrm{S}}  - d  e_{\Lambda} 
\label{dps2}
\end{equation}
The term in brackets is the differential of the full chemical potential Eq.~\ref{musfe}.  For evaluation of the remainder we return to the expression for the elastic energy density $e_{\Lambda}$ given in Eq.~\ref{epslamb}. The differential at fixed isochoric deformation $\mathbf{C}^i$ is proportional to the change in density. 
\begin{equation}
 d e_{\Lambda}   = \frac{1}{2} \left(\frac{ d \Lambda}{d \rho^{\mathrm{S}}} \right) \left( \mathrm{tr}  \, \mathbf{C}^i - 3 \right) d \rho^{\mathrm{S}} = \kappa_{\Lambda} e_{\Lambda} d \rho^{\mathrm{S}}
 \label{dps3}
 \end{equation}
 Inserted in Eq.~\ref{dps2} the $d \rho^{\mathrm{S}}$ terms cancel. The remaining term with prefactor  $\rho^{\mathrm{S}}$  leads to  Eq.~\ref{gibbsduhem}.
 
 For generalization of Eq.~\ref{dpbar0} to finite $\kappa_{\Lambda}$ we return to the  original pressure balance Eq.~\ref{dpbar}.
Expansion to first order in the stress induced density  increment   $\Delta \rho^{\mathrm{S}} = \rho^{\mathrm{S}} - \rho_{\mathrm{H}}$  gives
 \begin{equation}
 \rho_{\mathrm{H}} \Delta \bar{\mu}^{\mathrm{S}} + \left(  \left( \rho_{\mathrm{H}} + \Delta \rho^{\mathrm{S}}  \right) \kappa_{\Lambda} - 1 \right) e_{\Lambda} = 0
\label{dps4}
\end{equation}
 For the first term we have again exploited the Gibbs-Duhem relation for the short range pressure.   $\Delta \rho^{\mathrm{S}} e_{\lambda} $ is a second order term and is omitted in a first order approximation.  This changes Eq.~\ref{dps4} to
 \begin{equation}
 \rho_{\mathrm{H}} \left( \Delta \bar{\mu}^{\mathrm{S}} + \kappa_{\Lambda} e_{\Lambda} \right) -  e_{\Lambda} = 0
 \end{equation}
 The quantity in the brackets is the increment of the full chemical potential $\Delta \mu^{\mathrm{S}}$.  We are back where we were with the $\kappa_{\Lambda} = 0$ argument of section \ref{section:pomegap}.  Chemical equilibrium with the liquid prevents $\Delta \mu^{\mathrm{S}}$ to deviate from zero.  This  leaves us with $e_{\Lambda}= 0$.  Again finite shear elastic energy is ruled out by the combination of the balance equation for the chemical potential and thermodynamic pressure.  


\section{Evaluation Lam\'{e} moduli}\label{lame}

 The elastic response described by Eq.~\ref{deveps} is unusual. To see why,  we rewrite the linear stress-strain relation Eq.~\ref{deveps}  in regular  Lam\'{e} format\cite{lubarda20}
\begin{equation}
\mathbf{T} = 2 \mu \boldsymbol{\epsilon} + \lambda \left(\mathrm{tr} \boldsymbol{\epsilon} \right) \boldsymbol{1}
\label{Tlame}
\end{equation}
$\mu$ is normally referred to as the shear modulus.  The coefficient $\lambda$ seems to have no single commonly agreed name.  Comparison to Eq.~\ref{deveps} gives the expression of the linear Lam\'{e} moduli in terms of the  parameter $\Lambda$ multiplying the original non-linear stress (Eq.~\ref{devT}). 
\begin{equation}
\mu_{\Lambda} = \Lambda, \quad \lambda_{\Lambda} =  - \frac{2}{3} \Lambda 
\label{modulamb}
\end{equation}
As  intended the linear shear modulus $\mu_{\Lambda}$ is equal to $\Lambda$ (evaluated at the reference density).  $\lambda_{\Lambda}$ is also proportional to $\Lambda$ but with an abnormal negative sign.   Substituting in the Lam\'{e} equation for the bulk modulus\cite{lubarda20}
\begin{equation} 
K_{\Lambda} = \lambda_{\Lambda} + \frac{2}{3} \mu_{\Lambda} = 0
\label{klambda}
\end{equation}
reveals what is going on. The bulk modulus vanishes. The solid is infinitely compressible without the isotropic short range function $f^{\mathrm{S}}$ (see below).

The  Lam\'{e} moduli can be equally extracted from the quadratic energy density cast in the appropriate form\cite{lubarda20}
\begin{equation}
e = \mu \, \boldsymbol{\epsilon} : \boldsymbol{\epsilon} 
+ \tfrac{1}{2} \lambda \left( \mathrm{tr} \boldsymbol{\epsilon} \right)^2
\label{elame}
\end{equation}
Rearranging Eq.~\ref{epsquadr} accordingly we find
\begin{equation}
e_{\Lambda} = \Lambda \left(\epsilon_1^2 + \epsilon_2^2 + \epsilon_3^2 \right) 
 - \frac{\Lambda}{3} \left( \epsilon_1 + \epsilon_2 + \epsilon_3 \right)^2 
\end{equation}
confirming that the linear stress tensor Eq.~\ref{deveps} and Taylor expansion of the energy are consistent\cite{mistake}.

What still has to be accounted for is the effect of the short range forces which give rise to the pressure $\bar{p}^{\mathrm{S}}$ of Eq.~\ref{barps}.  This contribution is quantified by the  inverse of  a compressibility of Eq.~\ref{kapparho}. Pressures are  additive (see Eq.~\ref{psfe}) and therefore the net bulk modulus $K$ is the sum 
\begin{equation}
K = \frac{1}{\kappa_{\rho}} + K_{\Lambda} = \frac{1}{\kappa_{\rho}} 
\label{knet}
\end{equation}
where in the second equality we have used  Eq.~\ref{klambda}. The bulk modulus is determined by the short range forces only which are  purely hydrostatic in our scheme. Inserting Eqs.~\ref{modulamb} and \ref{knet} in Eq.~\ref{elame} the elastic energy of linear limit of the system defined by Eq.~\ref{liq2sol} can be written as
\begin{equation}
e = \Lambda  \, \boldsymbol{\epsilon} : \boldsymbol{\epsilon} 
+ \frac{1}{2} \left( \frac{1}{\kappa_{\rho}} - \frac{2}{3} \Lambda \right) \left( \mathrm{tr} \boldsymbol{\epsilon} \right)^2
\label{lamelamb}
\end{equation}
The ominous minus sign in Eq.~\ref{lamelamb} seems not to lead to instabilities.  In particular, the expression of  Young's modulus 
looks safe
\begin{equation}
E = \frac{\mu \left(3 \lambda +2 \mu \right)}{\lambda + \mu} = \frac{ 3 \Lambda }{1 + \kappa_{\rho} \Lambda/3}
\label{Ekappa}
\end{equation}
and familiar when written in standard form\cite{lubarda20}
\begin{equation}
E = \frac{ 9 \Lambda  K}{3K + \Lambda}
\label{Eyoung}
\end{equation}
The Poisson ratio then follows applying the appropriate relations for linear isotropic elasticity\cite{lubarda20}
\begin{equation}
\nu = \frac{3K -2 \Lambda}{2 \left(3K + \Lambda \right)}
\label{nupoisson}
\end{equation}
We also give the expression for  the compliances referred to in section \ref{section:gt}.
\begin{alignat}{2}
D_{iijj} & = \frac{1}{E} & \quad  & i = j 
\nonumber \\[4pt]
D_{iijj} & = -\frac{\nu}{E} &\quad  & i \neq j 
\label{compliance}
\end{alignat}

\section{Derivation GT pressure correction} \label{section:gtmath}
Assuming  the liquid pressure  correction is not too large $\Delta p^{\mathrm{L}}$  and  $\Delta \mu^{\mathrm{L}}$ are related by the isothermal Gibbs-Duhem equation and we can write $\Delta p^{\mathrm{L}}= \rho^{\mathrm{L}} \Delta \mu^{\mathrm{L}} $.  A similar Gibbs-Duhem relation also holds for atomic pressure and chemical potential of the solid: $\Delta \bar{p}^{\mathrm{S}}= \rho^{\mathrm{S}} \Delta \bar{\mu}^{\mathrm{S}} $.  This is simply a consequence of the definition Eq.~\ref{barps} of $\bar{p}^{\mathrm{S}}$.  
Substituting in Eq.~\ref{dpbargt} we find after rearranging 
\begin{equation}
\rho_{\mathrm{H}}^{\mathrm{S}} \Delta \bar{\mu}^{\mathrm{S}} +
 \rho^{\mathrm{S}} \kappa_{\Lambda}  e_{\Lambda}    
 -\rho_{\mathrm{H}}^{\mathrm{L}} \Delta \mu^{\mathrm{L}} = e_{\Lambda}
 \label{gt1}
\end{equation}
Next Eq.~\ref{dmubargt} is used to replace $\kappa_{\Lambda}  e_{\Lambda} $ giving
\begin{equation}
\rho_{\mathrm{H}}^{\mathrm{S}} \Delta \bar{\mu}^{\mathrm{S}} +
 \left( \rho_{\mathrm{H}}^{\mathrm{S}} + \Delta \rho^{\mathrm{S}} \right) 
  \left( \Delta \mu^{\mathrm{L}} - \Delta \bar{\mu}^{\mathrm{S}} \right)
 -\rho_{\mathrm{H}}^{\mathrm{L}} \Delta \mu^{\mathrm{L}} = e_{\Lambda}
 \label{gt2}
\end{equation}
We have also resolved the density $\rho^{\mathrm{S}}$ of the solid in a sum of the reference value $\rho_{\mathrm{H}}^{\mathrm{S}}$ in the initial hydrostatic state and the increment   $\Delta \rho^{\mathrm{S}}$ induced by the shear strain in the non-hydrostatic state.  The $\rho_{\mathrm{H}}^{\mathrm{S}} \Delta \bar{\mu}^{\mathrm{S}}$ terms cancel and Eq.~\ref{gt2} can be simplified to
\begin{equation}
\left( \rho_{\mathrm{H}}^{\mathrm{S}} - \rho_{\mathrm{H}}^{\mathrm{L}} \right)  \Delta \mu^{\mathrm{L}} +
 \Delta \rho^{\mathrm{S}} \left(  \Delta \mu^{\mathrm{L}} - \Delta \bar{\mu}^{\mathrm{S}} \right)
 = e_{\Lambda}
 \label{gt3}
\end{equation}
Reintroducing the chemostriction energy $\kappa_{\Lambda}  e_{\Lambda} $ using Eq.~\ref{dpbargt} (reversing the substitution in Eq.~\ref{gt1}) and gathering the terms proportional to $e_{\Lambda} $ we find
\begin{equation}
\left( \rho_{\mathrm{H}}^{\mathrm{S}} - \rho_{\mathrm{H}}^{\mathrm{L}} \right)  \Delta \mu^{\mathrm{L}} 
 = \left( 1 - \kappa_{\Lambda} \Delta \rho^{\mathrm{S}}  \right)  e_{\Lambda}
 \label{gt4}
\end{equation}
Finally changing back to specifying the state of the liquid reservoir in terms of pressure using once again Gibbs-Duhem we obtain Eq.~\ref{gteps}  as stated in the main text of  section \ref{section:gt}.

The other claim made in section \ref{section:gt} requiring proof is that Eq.~\ref{gteps0} obtained form Eq.~\ref{gteps} by setting $\kappa_{\Lambda} = 0$ can also directly 
derived from the Gibbs phase equilibrium condition Eq.~\ref{mulaccrv} discarding concerns about chemical equilibrium. As before,  we subtract the equilibrium equation at the hydrostatic reference state. This gives to first order in the difference of the relevant quantities
\begin{equation}
\frac{\Delta \psi^{\mathrm{S}} + \Delta p^{\mathrm{L}}}{\rho_{\mathrm{H}}^{\mathrm{S}}} 
+ \left( \psi_{\mathrm{H}}^{\mathrm{S}}  + p_{\mathrm{H}}^{\mathrm{L}} \right)  \Delta \left(\frac{1}{\rho^{\mathrm{S}}} \right)= \Delta  \mu^{\mathrm{L}}
\label{gt5}
\end{equation}
$\Delta \psi^{\mathrm{S}} = \psi^{\mathrm{S}} - \psi_{\mathrm{H}}^{\mathrm{S}}$ is the increase in Helmholtz free energy density.\cite{sekerka04}.  All other quantities have already been defined. Evaluating the change in inverse solid density to first order and rearranging we can write 
\begin{equation}
\Delta \psi^{\mathrm{S}} - \left( \psi_{\mathrm{H}}^{\mathrm{S}}  
+ p_{\mathrm{H}}^{\mathrm{L}} \right)  \left(\frac{ \Delta \rho^{\mathrm{S}}}{\rho_{\mathrm{H}}^{\mathrm{S}}} \right)= 
\left(\frac{\rho_{\mathrm{H}}^{\mathrm{S}} - \rho_{\mathrm{H}}^{\mathrm{L}}}{\rho_{\mathrm{H}}^{\mathrm{L}}} \right) \Delta p^{\mathrm{L}}
\label{gt6}
\end{equation}
where again the chemical potential and pressure of the liquid have been interchanged using the Gibbs-Duhem relation introducing a factor $1/\rho_{\mathrm{H}}^{\mathrm{L}}$

Next the Helmholtz free energy density change is resolved in the increment of the isotropic atomic energy  $f^{\mathrm{S}}$ and the shear strain energy. $e_{\Lambda} = 0$ in the hydrostatic reference but
 $f_{\mathrm{H}}^{\mathrm{S}}$ is finite and therefore 
\begin{equation}
\Delta \psi^{\mathrm{S}}  = \Delta f^{\mathrm{S}} + e_{\Lambda}
\end{equation}
or to first order in $\Delta \rho^{\mathrm{S}}$ recalling that $\bar{\mu}_{\mathrm{H}}^{\mathrm{S}} = \mu_{\mathrm{H}}^{\mathrm{S}}$
\begin{equation}
\Delta \psi^{\mathrm{S}}  = \mu_{\mathrm{H}}^{\mathrm{S}}  \Delta \rho^{\mathrm{S}} + e_{\Lambda}
\end{equation}
Substituting in Eq.~\ref{gt6} we now have
\begin{multline}
\Delta \psi^{\mathrm{S}} - \left( \psi_{\mathrm{H}}^{\mathrm{S}}  
+ p_{\mathrm{H}}^{\mathrm{L}} \right)  \left(\frac{ \Delta \rho^{\mathrm{S}}}{\rho_{\mathrm{H}}^{\mathrm{S}}} \right) =
\\
 \left(\mu_{\mathrm{H}}^{\mathrm{S}} \rho_{\mathrm{H}}^{\mathrm{S}} - f_{\mathrm{H}}^{\mathrm{S}} - p_{\mathrm{H}}^{\mathrm{L}} \right)
  \left(\frac{ \Delta \rho^{\mathrm{S}}}{\rho_{\mathrm{H}}^{\mathrm{S}}} \right)  + e_{\Lambda}
  \label{gt7}
 \end{multline}
The first term on the rhs vanishes because
\begin{equation}
\mu_{\mathrm{H}}^{\mathrm{S}} \rho_{\mathrm{H}}^{\mathrm{S}} - f_{\mathrm{H}}^{\mathrm{S}} - p_{\mathrm{H}}^{\mathrm{L}} =
  p_{\mathrm{H}}^{\mathrm{S}} - p_{\mathrm{H}}^{\mathrm{L}} = 0
\end{equation}
The hydrostatic pressures are balanced. As a result Eq.~\ref{gt7} is reduced to
\begin{equation}
\Delta \psi^{\mathrm{S}} - \left( \psi_{\mathrm{H}}^{\mathrm{S}}  
+ p_{\mathrm{H}}^{\mathrm{L}} \right)  \left(\frac{ \Delta \rho^{\mathrm{S}}}{\rho_{\mathrm{H}}^{\mathrm{S}}} \right) = e_{\Lambda}
 \end{equation}
Substituting for the lhs of Eq.~\ref{gt6} yields
\begin{equation}
 e_{\Lambda} = 
\left(\frac{\rho_{\mathrm{H}}^{\mathrm{S}} - \rho_{\mathrm{H}}^{\mathrm{L}}}{\rho_{\mathrm{H}}^{\mathrm{L}}} \right) \Delta p^{\mathrm{L}}
\label{gtend}
\end{equation}
which is equivalent  to Eq.~\ref{gteps0}. As far as we can see,  the lengthy sequence of substitutions and cancellations  leading to  Eq.~\ref{gtend} unrolls the highly compact two line derivation given by Nozi\'{e}res in Refs.~\citenum{nozieres93} and \citenum{nozieres95}. It was certainly intended to be so.

\section*{References}


%

\end{document}